\documentclass[longauth]{aa} % for the long lists of affiliations 
\usepackage{graphicx}
\usepackage{txfonts}
\usepackage{natbib}
\usepackage{longtable}
\def\pdeg{\ifmmode $\setbox0=\hbox{$^{\circ}$}\rlap{\hskip.11\wd0 .}$^{\circ}
  \else \setbox0=\hbox{$^{\circ}$}\rlap{\hskip.11\wd0 .}$^{\circ}$\fi}

\begin{document}

\title{Fragmentation and kinematics in high-mass star formation}

\subtitle{CORE-extension targeting two very young high-mass star-forming regions}

   \author{H.~Beuther
          \inst{1}
          \and
          C.~Gieser
          \inst{1}
          \and
          S.~Suri
          \inst{1}
          \and
          H.~Linz
          \inst{1}
         \and
          P.~Klaassen
          \inst{2}
         \and
          D.~Semenov
          \inst{1}
         \and
          J.M.~Winters
          \inst{3}
         \and
          Th.~Henning
          \inst{1}
         \and
          J.\,D.~Soler
          \inst{1}
         \and
          J.S.~Urquhart
          \inst{4}
         \and
          J.~Syed
          \inst{1}
         \and
          S.~Feng
          \inst{5}
         \and
          T.~M\"oller
          \inst{6}
         \and
          M.\ T.\ Beltr\'an 
          \inst{7}
         \and
          \'A. S\'anchez-Monge
          \inst{6}
         \and
          S.N.~Longmore
          \inst{8}
         \and
          T.~Peters
          \inst{9}
         \and
          J.~Ballesteros-Paredes
          \inst{10}
         \and
          P.~Schilke
          \inst{6}
         \and
          L.~Moscadelli
          \inst{7}
         \and
          A.~Palau
          \inst{10}
         \and
          R.~Cesaroni
          \inst{7}
         \and
          S.~Lumsden
          \inst{11}
         \and
          R.~Pudritz
          \inst{12}
         \and
          F.~Wyrowski
          \inst{13}
         \and
          R.~Kuiper
          \inst{14}
         \and
          A.~Ahmadi
          \inst{15} 
}
   \institute{$^1$ Max Planck Institute for Astronomy, K\"onigstuhl 17,
     69117 Heidelberg, Germany, \email{name@mpia.de}\\
     $^2$ UK Astronomy Technology Centre, Royal Observatory Edinburgh, Blackford Hill, Edinburgh EH9 3HJ, UK\\
     $^3$ IRAM, 300 rue de la Piscine, Domaine Universitaire de Grenoble, 38406 St.-Martin-d’Hères, France\\
     $^4$ Centre for Astrophysics and Planetary Science, University of Kent, Canterbury, CT2 7NH, UK\\
     $^5$ Chinese Academy of Sciences Key Laboratory of FAST, National Astronomical Observatory of China, Datun Road 20, Chaoyang, Bejing, 100012, P.~R.~China\\
     $^6$ I.~Physikalisches Institut, Universit\"at zu K\"oln, Z\"ulpicher Str. 77, 50937 K\"oln, Germany\\
     $^7$ INAF, Osservatorio Astrofisico di Arcetri, Largo E. Fermi 5, I-50125 Firenze, Italy\\
     $^8$ Astrophysics Research Institute, Liverpool John Moores University, 146 Brownlow Hill, Liverpool L3 5RF, UK\\
     $^9$  Max-Planck-Institut für Astrophysik, Karl-Schwarzschild-Str. 1, D-85748 Garching, Germany\\
     $^{10}$ Instituto de Radioastronomıa y Astrofısica, Universidad Nacional Autonoma de Mexico, PO Box 3-72, 58090 Morelia, Michoacan, Mexico\\
     $^{11}$ School of Physics \& Astronomy, E.C. Stoner Building, The University of Leeds, Leeds LS2 9JT, UK\\
     $^{12}$ Department of Physics and Astronomy, McMaster University, 1280 Main St. W, Hamilton, ON L8S 4M1, Canada\\
     $^{13}$ Max-Planck-Institut für Radioastronomie, Auf dem Hügel 69, 53121 Bonn, Germany\\
     $^{14}$ Institute of Astronomy and Astrophysics, University of T\"ubingen, Auf der Morgenstelle 10, 72076, T\"ubingen, Germany\\
     $^{15}$ Leiden Observatory, Leiden University, PO Box 9513, 2300 RA Leiden, The Netherlands
}

   \date{Version of \today}

%   \abstract{}
% \abstract{}{}{}{}{} 
% 5 {} token are mandatory  
\abstract
  % context heading (optional)
% {} leave it empty if necessary
    {The formation of high-mass star-forming regions from their
      parental gas cloud and the subsequent fragmentation processes lie
      at the heart of star formation research.}
% aims heading (mandatory)
    {We aim to study the dynamical and fragmentation properties at very
      early evolutionary stages of high-mass star formation.}
% methods heading (mandatory)
    {Employing the NOrthern Extended Millimeter Array (NOEMA) and the
      IRAM 30\,m telescope, we observed two young high-mass star-forming
      regions, ISOSS22478 and ISOSS23053, in the 1.3\,mm continuum and
      spectral line emission at a high angular resolution
      ($\sim$0.8$''$).}
% results heading (mandatory)  
{We resolved 29 cores that are mostly located along filament-like
  structures. Depending on the temperature assumption, these cores
  follow a mass-size relation of approximately $M\propto r^{2.0\pm
    0.3}$, corresponding to constant mean column densities. However,
  with different temperature assumptions, a steeper mass-size relation
  up to $M\propto r^{3.0\pm 0.2}$, which would be more likely to correspond to
  constant mean volume densities, cannot be ruled out. The correlation
  of the core masses with their nearest neighbor separations is
  consistent with thermal Jeans fragmentation.  We found hardly any core
  separations at the spatial resolution limit,  indicating that the data  resolve the large-scale fragmentation
  well. Although the kinematics of the two regions appear very
  different at first sight -- multiple velocity components along
  filaments in ISOSS22478 versus a steep velocity gradient of more
  than 50\,km\,s$^{-1}$\,pc$^{-1}$ in ISOSS23053 -- the findings can
  all be explained within the framework of a dynamical cloud collapse
  scenario.}
% conclusions heading (optional), leave it empty if necessary
{While our data are consistent with a dynamical cloud collapse
  scenario and subsequent thermal Jeans fragmentation, the importance
  of additional environmental properties, such as the magnetization of
  the gas or external shocks triggering converging gas flows, is
  nonetheless not as well constrained and would require future investigation.}


\titlerunning{Kinematics of cloud collapse during high-mass star
  formation}

\maketitle
 
\begin{figure*}[htb]
\includegraphics[width=0.48\textwidth]{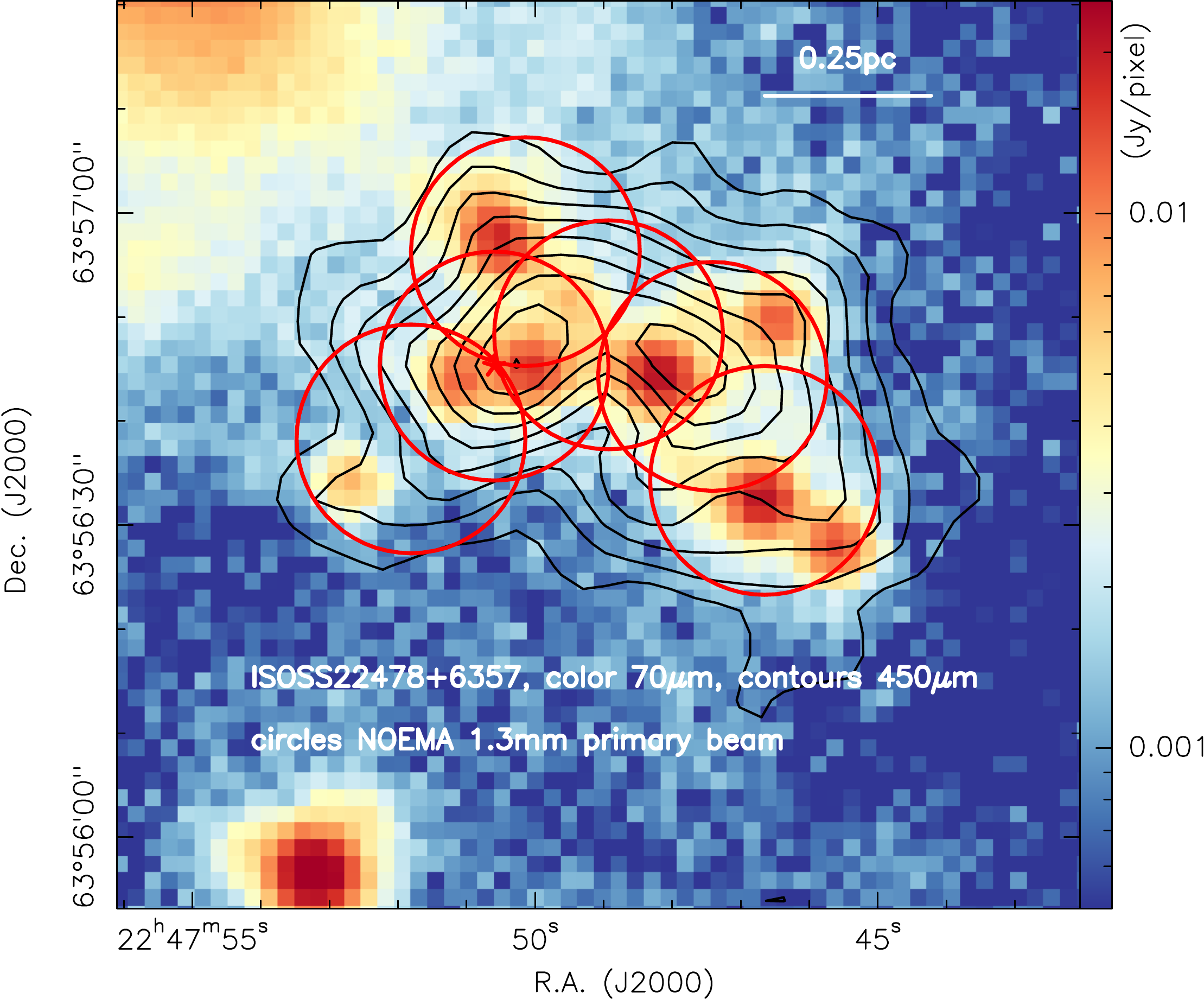}
\includegraphics[width=0.505\textwidth]{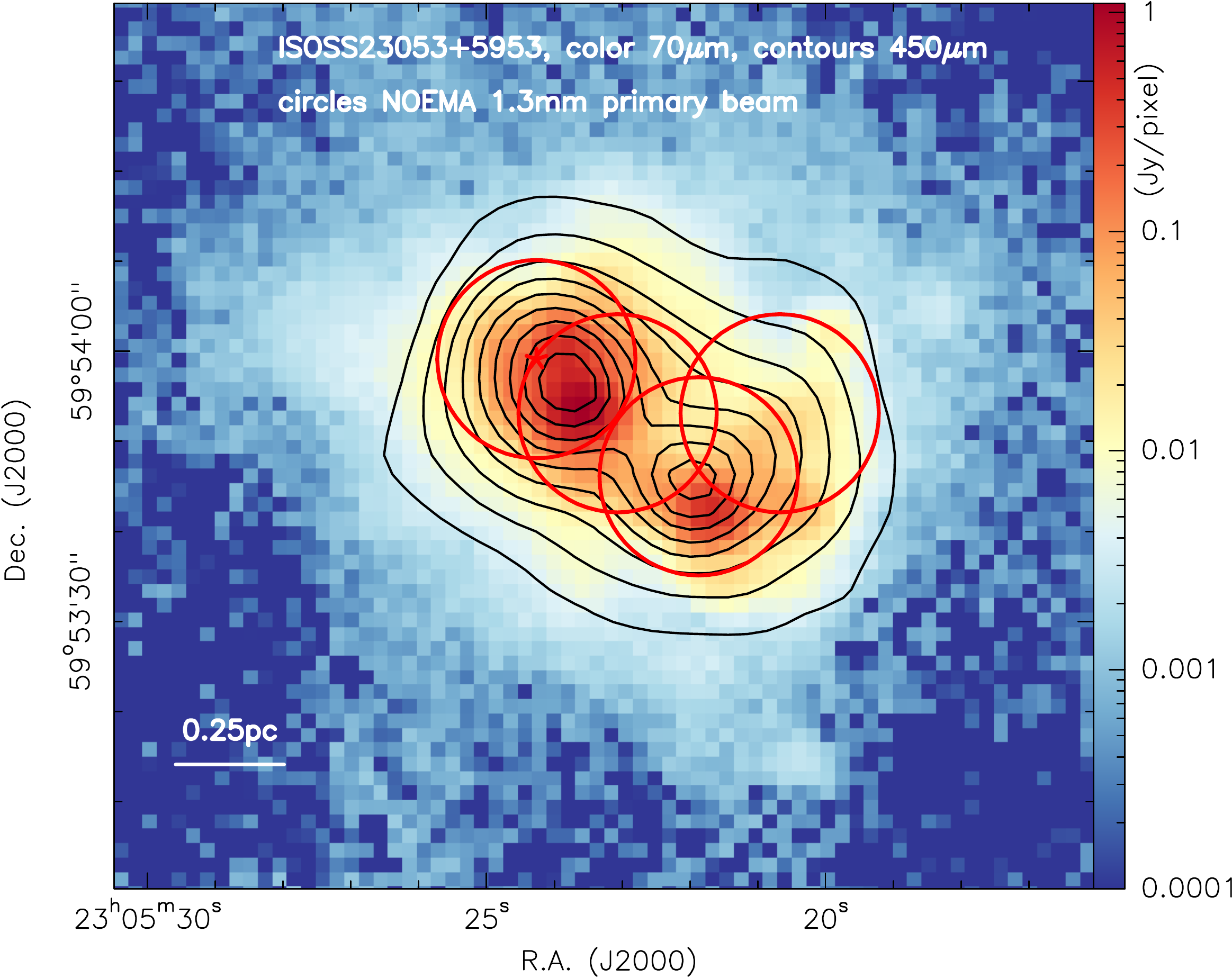}
\caption{Overview of the ISOSS22478+6357 (left) and ISOSS23053 (right)
  regions. The color scales and contours show the 70 and 450\,$\mu$m
  emission from Herschel and SCUBA, respectively
  \citep{ragan2012b,difrancesco2008}. The contour levels are from 18
  to 98\% of the peak emission of 1.88\,Jy\,beam$^{-1}$ and
  14.88\,Jy\,beam$^{-1}$ for the two regions, respectively. Linear
  scale bars are shown and the red circles outline the NOEMA mosaics
  with a primary beam FWHM of $22''$.}
\label{overview} 
\end{figure*} 

\section{Introduction}
\label{intro}

During the hierarchical process of star formation, the gas has to flow
from the large cloud scales down to the smallest scales, where
disks and protostars form. While questions regarding the small-scale
structure and the presence of disks toward more evolved high-mass
protostellar objects have been subject to intense study for several
decades (e.g., \citealt{cesaroni2007,beltran2016,beuther2018b}), the
earliest evolutionary stages of collapse and fragmentation during the
formation processes of high-mass stars have not been
investigated in depth thus far. This is in part related to the short evolutionary
time-scales of early massive star formation (e.g.,
\citealt{beuther2006b,zinnecker2007, tan2014}) and the fact that
hardly any high-mass starless core candidates manage to survive further scrutiny
when additional star formation indicators become detected (e.g.,
\citealt{motte2007,tan2016,feng2016b}). A few exceptional regions
exist that have remained starless even with high-resolution millimeter
interferometric observations (e.g., G11.92
\citealt{cyganowski2014,cyganowski2017}; IRDC18310-4
\citealt{beuther2013a,beuther2015a}; W43-MM1
\citealt{nony2018}). However, at least the latter two examples are at
comparably large distances of 3.4 and 4.9\,kpc, respectively, reducing
the ability to investigate structures at small physical spatial
scales. Furthermore, recent investigations with the Atacama Large
Millimeter Array (ALMA) and the Submillimeter Array (SMA) also started
to further dissect samples of very young high-mass star-forming
regions (e.g., \citealt{sanhueza2019,svoboda2019,li2019,li2020b}).

Important questions regarding the earliest evolutionary stages of
high-mass star formation are related to the fragmentation properties
of the dense gas and the associated kinematic properties. For example,
we consider whether these very young regions fragment early on into many lower-mass
cores that may then competitively accrete gas from the surrounding
envelope (e.g., \citealt{bonnell2007,smith2009a}). Alternatively,  they may fragment
into more massive cores with potentially less low-mass cores early on
(e.g., \citealt{tan2014,zhang2015,csengeri2017,beuther2018b})? We ask how
important filamentary accretion processes are (e.g.,
\citealt{schneider2010,peretto2014,andre2014,chira2018,hennebelle2018,padoan2020}) and what the relative importance is of cloud-scale gravo-turbulent or
thermal fragmentation versus fragmentation of the gravitationally
unstable accretion flows for the star formation process (e.g.,
\citealt{peters2010b})? Furthermore, it has been argued that high-mass
star formation starts and proceeds in turbulent gas clumps (e.g.,
\citealt{mckee2003}). However, recent high-spatial-resolution studies
of a few very young high-mass star-forming have regions revealed that at
high enough spatial and spectral resolution, the spectral lines resolve
into multiple components, each with line widths that are nearly consistent with
thermal line broadening (e.g.,
\citealt{beuther2015a,hacar2018,li2020}). Therefore, it may well be
the case that the individual cores do not have turbulent properties but it is, rather,
only the superposition of many of these individual cores and spectral
components mimicking the broader lines that are typically seen with single-dish
telescopes at lower spatial resolution (e.g.,
\citealt{smith2013,hacar2017b}). At a somewhat coarser resolution of
$>$10\,000\,au, similar results indicating sub-sonic non-thermal
motions were also found by \citet{ragan2015} and \citet{sokolov2018}
in lower-density lines of N$_2$H$^+$ and NH$_3$, respectively.

With the goal of studying the kinematic and fragmentation properties at
the onset of high-mass star formation, we employ high spatial
resolution and high-density tracer observations at mm wavelengths
toward two regions at very early evolutionary stages. The targets are
part of the IRAM NOEMA large program CORE aimed at studying the
fragmentation, disk properties, kinematics, and chemistry during the
formation of high-mass stars \citep{beuther2018b}. In the following,
we summarize the main aspects of the CORE program.

\section{The CORE and CORE-extension projects}

With the goal to study the fragmentation, disk, outflow, and chemical
properties of high-mass star formation, we have embarked on the NOEMA
large program CORE, which observes 20 high-mass star-forming regions at
a high spatial resolution ($0.3''-0.4''$) in the 1.3\,mm line and
continuum emission. The sample was selected based on being mainly in the
evolutionary stage of high-mass protostellar objects (HMPOs),
also known as MYSOs (massive young stellar objects). An overview and
more details about the program can be found in
\citet{beuther2018b}\footnote{See also http://www.mpia.de/core}.

To broaden the scope of CORE to even younger evolutionary stages, we
conducted the CORE-extension part of the project, observing two much
younger regions (ISOSS22478 and ISOSS23053) also at high spatial
resolution and with a similar spectral setup. The goal of this
CORE-extension is to investigate earlier evolutionary stages with a
particular focus on the gas dynamics and chemical properties at the
onset of high-mass star formation. Selection criteria for their
  youth were mainly their low temperatures and luminosities as well as the low
  luminosity-to-mass ratios $L/M$ ($\sim$0.4 and $\sim$2.2 for
  ISOSS22478 and ISOSS2305, respectively). These $L/M$ ratios are
  typically more than an order of magnitude below those of the
  original CORE sample \citep{beuther2018b}. Below, we introduce the
target regions and Sect. \ref{core-extension} presents the technical
details of the CORE-extension project.

\begin{figure*}[htb]
\includegraphics[width=0.5\textwidth]{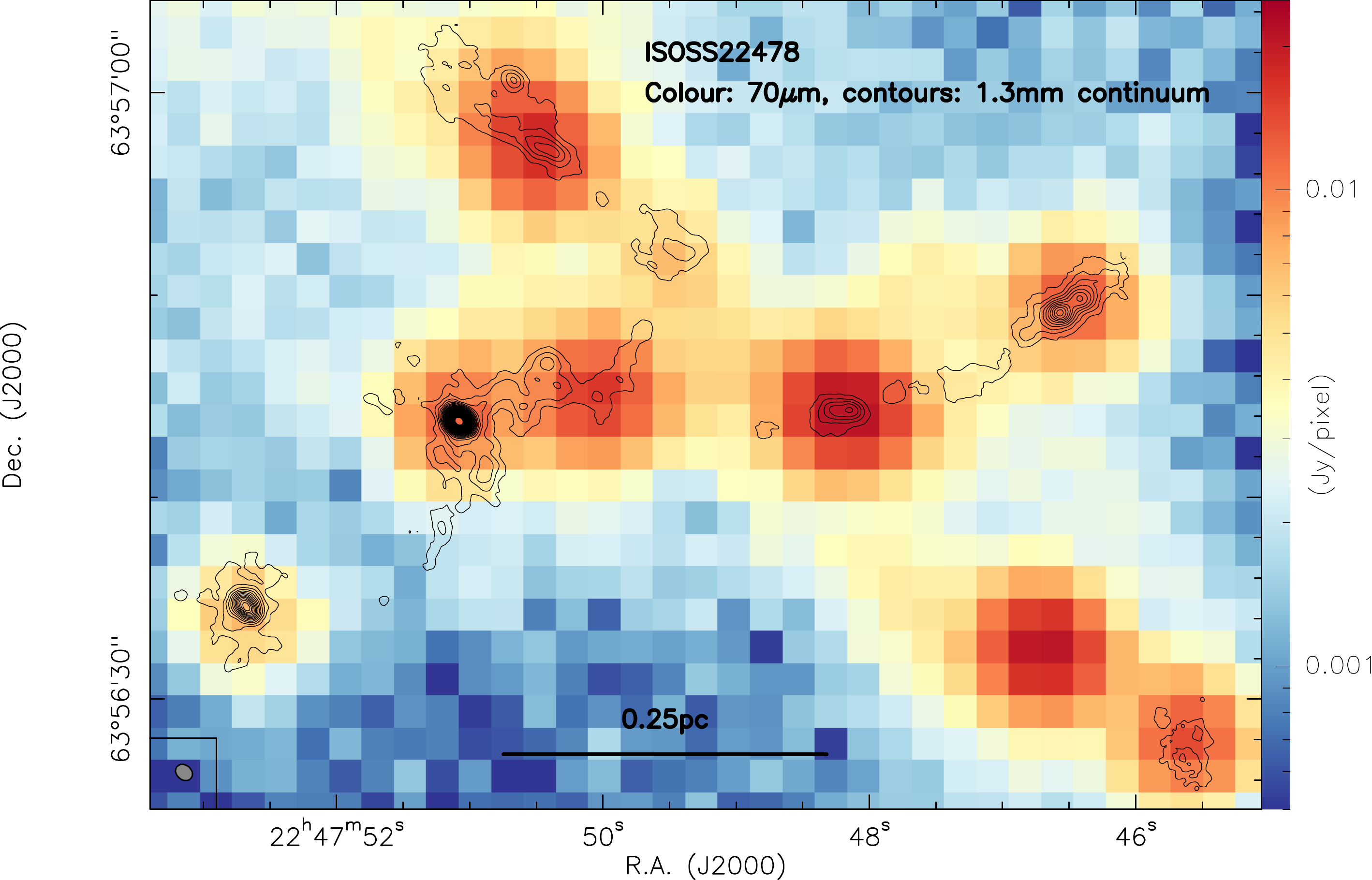}
\includegraphics[width=0.48\textwidth]{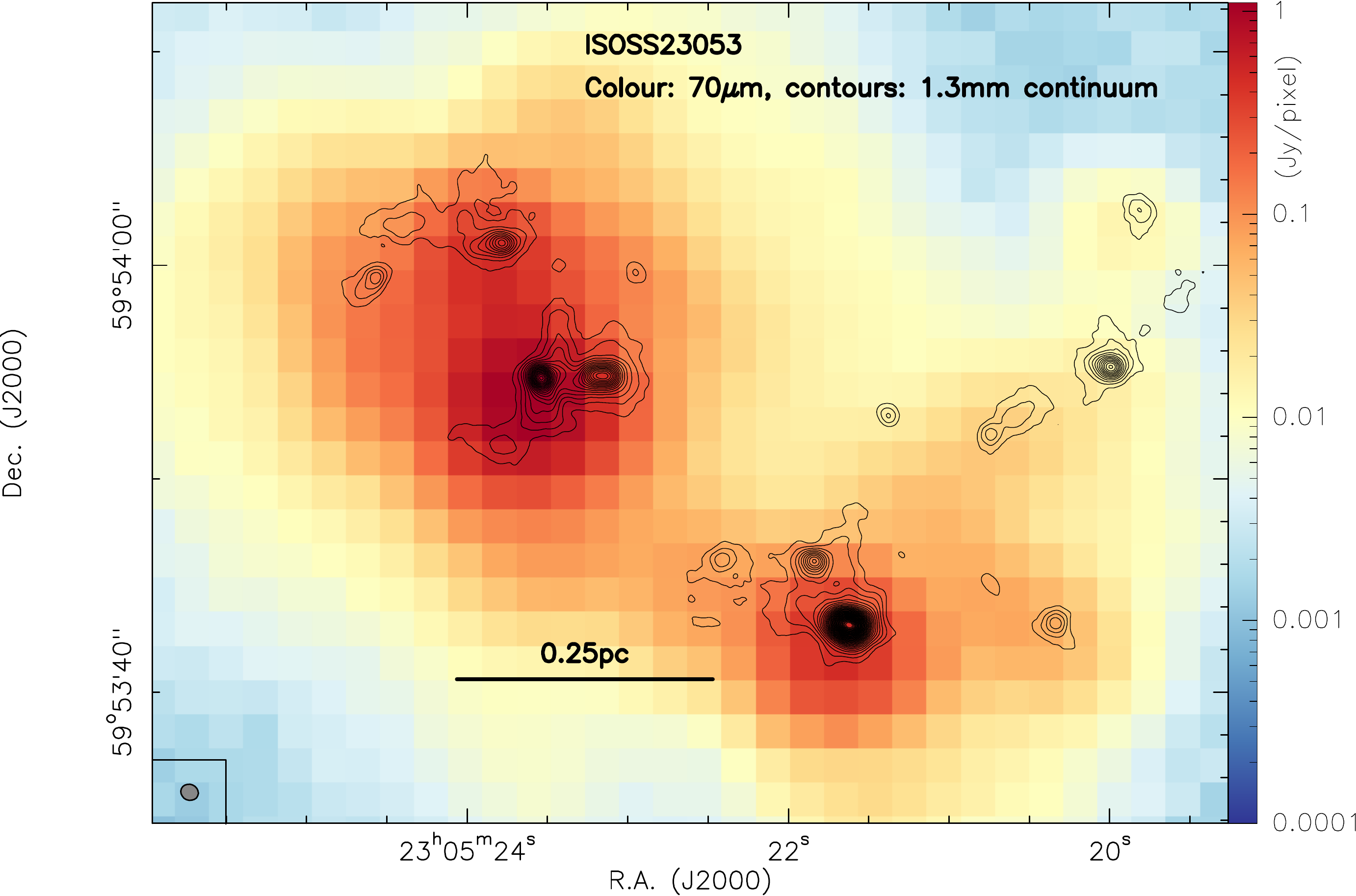}
\caption{NOEMA 1.3\,mm continuum data for ISOSS22478 (left) and
  ISOSS23053 (right). The color scales show the 70\,$\mu$m Herschel
  data and the contours present the 1.3\,mm NOEMA continuum data. The
  contour levels are in steps of $3\sigma$ with the $1\sigma$ level of
  57\,$\mu$Jy\,beam$^{-1}$ for ISOSS22478 and 160\,$\mu$Jy\,beam$^{-1}$ for
  ISOSS23053. Linear scale bars and the 1.3\,mm beam size are shown.}
\label{cont} 
\end{figure*}

\subsection{Target regions}
\label{targets}

While the main part of the program targets 20 HMPOs with already
embedded massive young stellar objects, we also observed two younger
regions in slightly more extended mosaics to investigate the earliest
formation processes. These two regions were originally identified via
the ISO Serendipity Survey (ISOSS), which observed the sky
serendipitously also during slew times with the Infrared Space
Observatory (ISO) at 170\,$\mu$m \citep{bogun1996,krause2004}. Since
the regions are located in the outer Galaxy, and not in front of strong
Milky Way background emission, they do not appear as infrared dark clouds
(IRDCs). However, their general physical properties put them in a
similarly young evolutionary stage as typical IRDCs.

{\bf ISOSS22478+6357:} Earlier studies of this region were conducted
by \citet{hennemann2008}, \citet{ragan2012b} and \citet{bihr2015}.  At
a distance of $\sim$3.23\,kpc, the region has a mass reservoir of
$\sim$140\,M$_{\odot}$, placing it in the category of
intermediate-mass star-forming regions. The average temperatures based
on NH$_3$ and dust observations are between $\sim$13\,K and
$\sim$21\,K, and the region still has a low total luminosity of
$\sim$55\,L$_{\odot}$ \citep{ragan2012b,bihr2015}. Figure
\ref{overview} (left panel) gives a large-scale overview of the region
comprising the Herschel 70\,$\mu$m emission and the SCUBA 450\,$\mu$m
data \citep{ragan2012b,difrancesco2008}.

{\bf ISOSS23053+5953:} This region has also previously been
investigated, for example, by \citet{birkmann2007}, \citet{ragan2012b}
and \citet{bihr2015}. At a distance of $\sim$4.31\,kpc, the region has
a mass reservoir and luminosity of 610\,M$_{\odot}$ and
1313\,L$_{\odot}$, respectively.  Average NH$_3$ and dust temperature
for the region are $\sim$18\,K and $\sim$22\,K
\citep{bihr2015,ragan2012b}. Figure \ref{overview} (right panel) shows
the corresponding large-scale overview image of the region. While
early studies already showed signatures of infall
\citep{birkmann2007}, an analysis of the kinematics of this region
based on VLA (Very Large Array) NH$_3$ observations revealed
indications of two velocity components that may collide at the
positions of the central cores \citep{bihr2015}.

\begin{table*}[htb]
  \caption{Observational parameters.}
  \label{lines}
  \begin{center}
  \begin{tabular}{lrrrrr}
    \hline \hline
    & & \multicolumn{2}{c}{ISOSS22478} & \multicolumn{2}{c}{ISOSS23053} \\
    setup & freq. & $1\sigma$ & beam& $1\sigma$ & beam  \\
    & (GHz) & (mJy\,beam$^{-1}$) & ($''$) & (mJy\,beam$^{-1}$) & ($''$)\\
    \hline
continuum & 224.970 & 0.057 & $0.92\times 0.73$ & 0.16 & $0.84\times 0.74$\\
DCO$^+$(3--2) & 216.113 & 6.0 & $0.97\times 0.77$ & 6.0 & $0.91\times 0.78$ \\
SiO(5--4)    & 217.105 & 7.5 & $0.97\times 0.77$ & 6.0 & $0.89\times 0.78$\\
H$_2$CO$(3_{0,3}-2_{0,2})$ & 218.222 & 7.0 & $0.97\times 0.77$ & 6.0 & $0.89\times 0.77$\\
H$_2$CO$(3_{2,2}-2_{2,1})$ & 218.476 & 7.5 & $0.97\times 0.77$ & 6.0 & $0.89\times 0.77$\\
H$_2$CO$(3_{2,1}-2_{2,0})$ & 218.760 & 7.0 & $0.97\times 0.77$ & 6.0 & $0.89\times 0.77$\\
    \hline \hline
  \end{tabular}
  ~\\
  Note: The $1\sigma$ line rms corresponds to 0.3\,km\,s$^{-1}$ velocity resolution.
  \end{center}
\end{table*}

\subsection{The CORE-extension technical setup}
\label{core-extension}

While the original CORE program utilized the old receiver and backend
system with only a $\sim$4\,GHz wide band in a single sideband between
217.167 and 220.834\,GHz, the upgraded NOEMA now allows us to use, in
dual-polarization, a much broader bandwidth in the double-sideband mode
with 7.8\,GHz in each sideband. Specifically, we are covering the
frequency ranges from $\sim$213.3 to $\sim$221.1\,GHz as well as from
$\sim$228.8 to $\sim$236.6\,GHz. This spectral setup covers a
multitude of strong spectral lines from many important molecules,
for instance, deuterated species such as DCO$^+$, DNC, or N$_2$D$^+$, the
temperature tracers H$_2$CO or CH$_3$CN, CO, and its isotopologues
$^{13}$CO and C$^{18}$O, shock tracers like SiO or larger molecules
such as CH$_3$OH or HCCCH. While the entire bandpass is covered at low
spectral resolution (2\,MHz corresponding to $\sim$2.7\,km\,s$^{-1}$
at 220\,GHz), all important lines are also covered by separate
high-spectral-resolution units (0.06\,MHz corresponding to
$\sim$0.08\,km\,s$^{-1}$ at 220\,GHz).

More details about the spectral setup and the chemical properties of
the regions will be given in a future paper by Gieser et al.~(in
prep.). Here, we focus on the dynamical properties and
potential gas flows in these two regions.

\section{Observations and data}
\label{data}

The two regions were observed with NOEMA in February and March 2019
with ten antennas in the A, C, and D configurations covering baseline
lengths roughly between  18 and 774\,m. The sources ISOSS22478 and
ISOSS23053 were observed as six-field and four-field mosaics, respectively
(see red circles in Fig.~\ref{overview}). The phase centers and
velocities of rest $v_{\rm lsr}$ were R.A.~(J2000.)  22:47:49.22994,
Dec.~63:56:45.2796 (Gal.~long./lat.~ 109.86/4.26\,degs) and $v_{\rm
  lsr}=-39.7$\,km\,s$^{-1}$ for ISOSS22478 and R.A.~(J2000.)
23:05:22.46953, Dec.~59:53:52.6192 (Gal.~long./lat.~
  109.99/-0.28\,degs) and $v_{\rm lsr}=-51.7$\,km\,s$^{-1}$ for
ISOSS23053. Flux and bandpass calibration were conducted for both
regions with MWC349 and 3c454.3, respectively. Regularly interleaved
observations of nearby quasars were used for phase and amplitude
calibration, namely, 0016+731 for ISOSS22078 and J2223+628
for ISOSS23053.

\begin{figure*}[htb]
\includegraphics[width=0.50\textwidth]{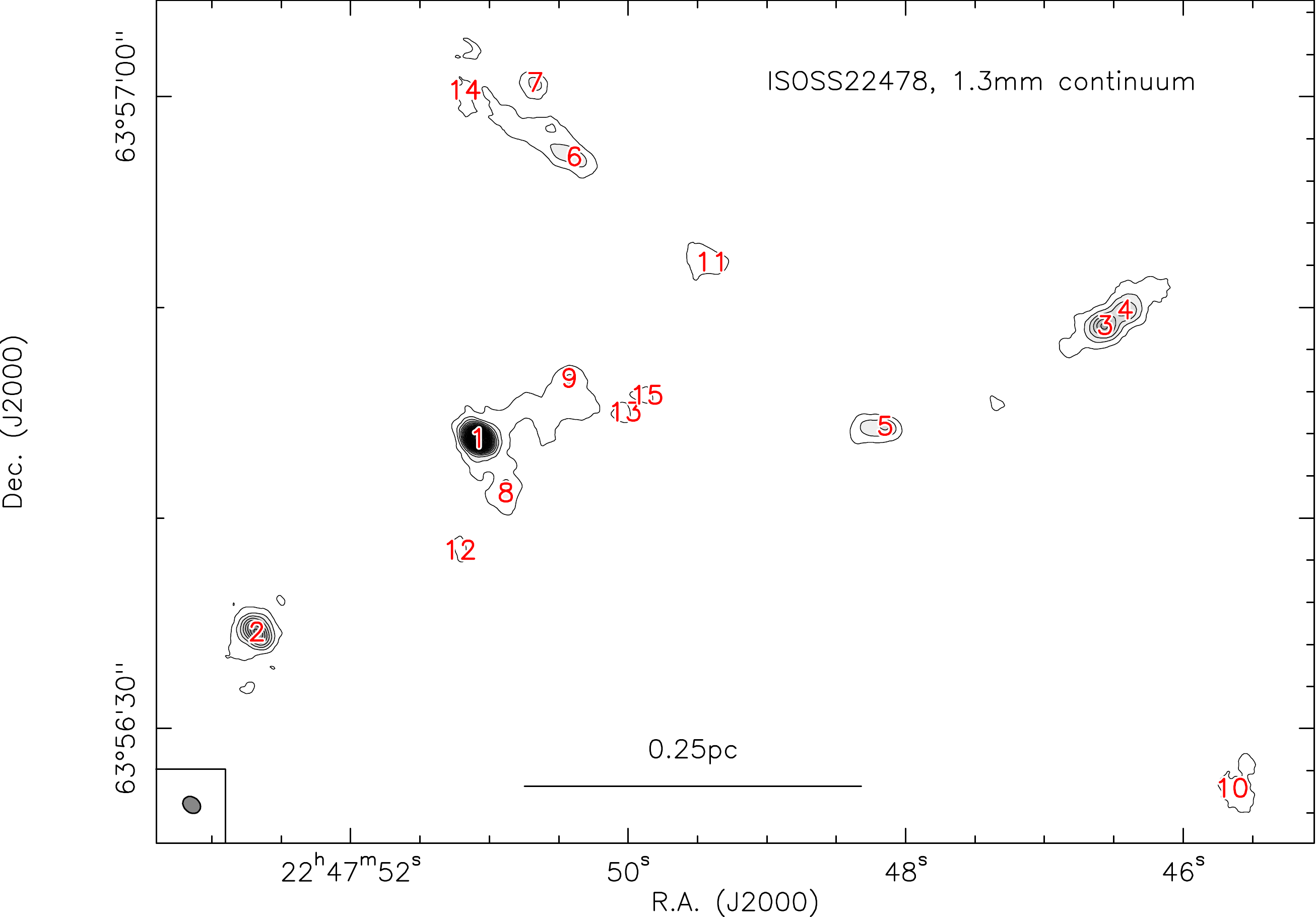}
\includegraphics[width=0.48\textwidth]{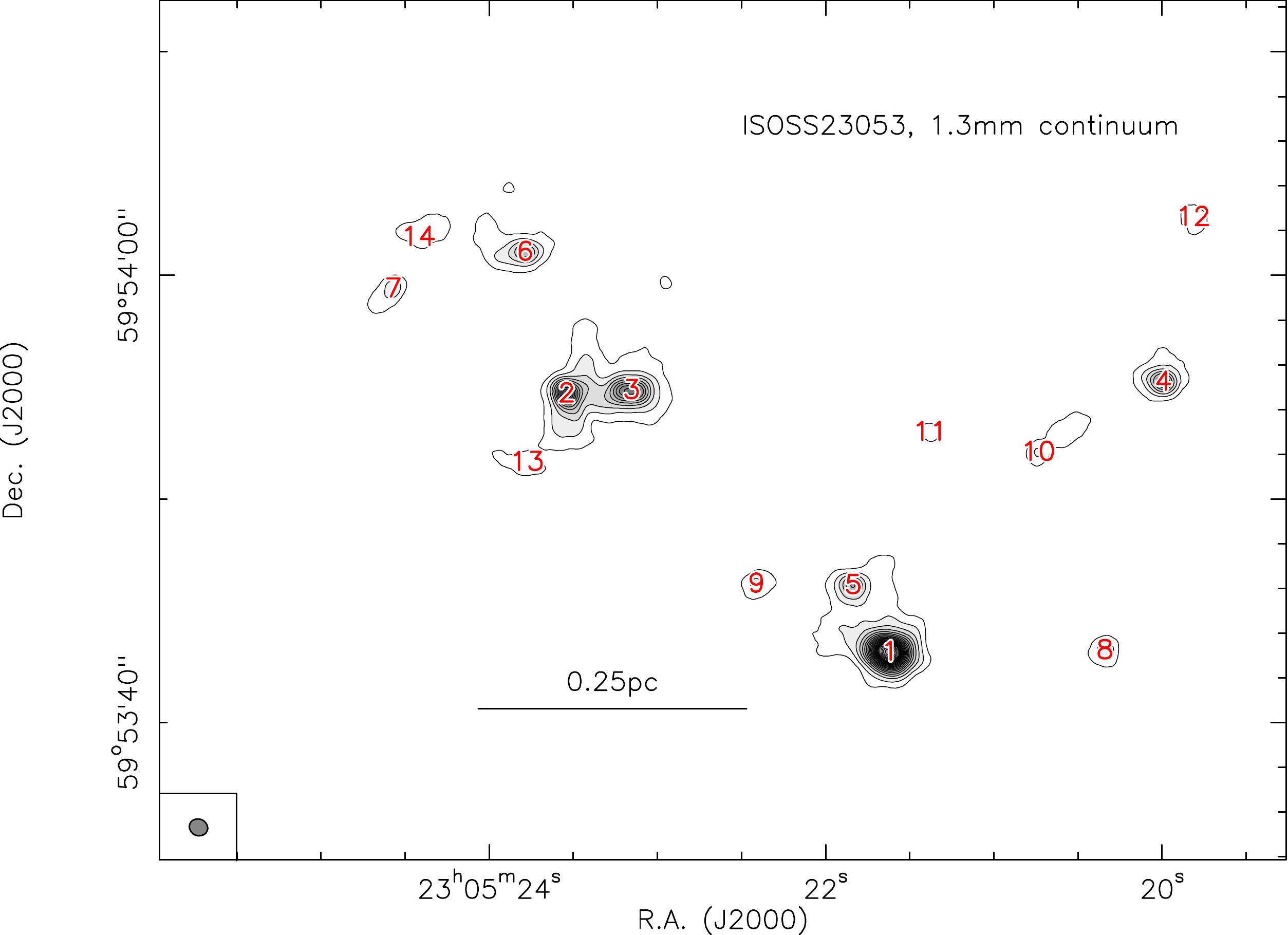}
\caption{NOEMA 1.3\,mm continuum data for ISOSS22478 and
  ISOSS23053. The grey-scale and contours are now done in $5\sigma$
  steps of 0.238 and 0.8\,mJy\,beam$^{-1}$ for ISOSS22478 and
  ISOSS23053, respectively. The red numbers label the identified
  cores. A linear scale bar and the 1.3\,mm beam size are shown.}
\label{cont_num} 
\end{figure*}

Calibration and imaging of the data was performed with the {\sc clic}
and {\sc mapping} software of the {\sc gildas}
package\footnote{http://www.iram.fr/IRAMFR/GILDAS}. To achieve optimal imaging
quality for these mosaics, we applied natural weighting during the
imaging process. To create the continuum data, only the line-free
parts of the spectrum were used. Since the regions are comparably
line-poor, only small parts of the entire bandpass (Sect.
\ref{core-extension}) had to be excluded. Specifically, we excluded
280\,MHz and 1800\,MHz of the entire bandpass of $\sim$15\,600\,MHz
for ISOSS22478 and ISOSS23053, respectively. For the 1.3\,mm continuum
data, this resulted in synthesized beams of $0.92''\times 0.73''$
(P.A.~51\,deg) and $0.84''\times 0.74''$ (P.A.~67\,deg) for ISOSS22478
and ISOSS23053, respectively. This corresponds to approximate linear
resolution elements of $\sim$2700 and $\sim$3500\,au,
respectively. The corresponding $1\sigma$ continuum rms values are
0.057\,mJy\,beam$^{-1}$ and 0.16\,mJy\,beam$^{-1}$, respectively.  The
absolute flux scale is estimated to be correct within 20\%.

For the spectral line data, we obtained complementary single-dish
observations with the IRAM 30\,m telescope to compensate for the
missing short spacings. These 30\,m observations were conducted in the
on-the-fly mode typically achieving rms values of $\sim$0.1\,K
($T_{\rm{mb}}$).

The NOEMA and 30\,m data were then combined in the imaging process
with the task {\sc uv\_short}, and we
present the merged data for DCO$^+$(3--2), SiO(5--4) and several
H$_2$CO lines at a spectral resolution of 0.3\,km\,s$^{-1}$. Again,
natural weighting was applied, and the final $1\sigma$ rms and
synthesized beam values are presented in Table \ref{lines}. The other
spectral lines and a detailed chemical analysis of the regions will be
presented in Gieser et al.~(in prep.).

\begin{table*}[htb]
  \caption{Continuum parameters.}
  \label{cont_par}
  \begin{center}
  \begin{tabular}{lrrrrrrrrrrrr}
    \hline \hline
    \# & R.A. & Dec. & $\Delta x$ & $\Delta y$ & $S_{\rm{peak}}$ & $S$ & $r$ & $M_{\rm 20K}$ & $N_{\rm peak, 20K}$ & $T_{\rm H2CO}$ & $M_{\rm T_{\rm H2CO}}$ & $N_{\rm peak, T_{\rm H2CO}}$   \\
       & (J2000.0) & (J2000.0) & ($''$)     & ($''$)     & $\left(\frac{\rm mJy}{\rm beam}\right)$ & (mJy) & (au) & (M$_{\odot}$) & ($10^{23}$cm$^{-2}$) & (K) & (M$_{\odot}$) & ($10^{23}$cm$^{-2}$) \\
    \hline
    \multicolumn{10}{c}{ISOSS22478} \\
    \hline
  1  & 22:47:51.08 & 63:56:43.7 &  12.20  &  -1.54  & 7.690  & 12.585  & 5007 & 4.45  & 6.63 & 42  & 1.84 & 2.743 \\  
  2  & 22:47:52.68 & 63:56:34.6 &  22.71  & -10.73  & 2.380  &  5.136  & 3967 & 1.82  & 2.05 & 13  & 3.33 & 3.753 \\  
  3  & 22:47:46.56 & 63:56:49.1 & -17.57  &   3.82  & 1.800  &  4.020  & 3841 & 1.42  & 1.55 & 33  & 0.78 & 0.853 \\  
  4  & 22:47:46.42 & 63:56:49.8 & -18.52  &   4.56  & 1.250  &  3.488  & 3915 & 1.23  & 1.08 & 22  & 1.10 & 0.963 \\  
  5  & 22:47:48.15 & 63:56:44.3 &  -7.13  &  -0.96  & 0.890  &  2.015  & 3108 & 0.71  & 0.77 & 37  & 0.34 & 0.373 \\  
  6  & 22:47:50.39 & 63:56:57.1 &   7.64  &  11.83  & 0.780  &  4.081  & 4986 & 1.44  & 0.67 & 27  & 0.97 & 0.453 \\  
  7  & 22:47:50.67 & 63:57:00.6 &   9.48  &  15.36  & 0.730  &  0.787  & 2096 & 0.28  & 0.63 & 10$^*$  & 0.76 & 1.713 \\  
  8  & 22:47:50.88 & 63:56:41.2 &  10.88  &  -4.12  & 0.680  &  1.895  & 3326 & 0.67  & 0.59 & 27  & 0.47 & 0.413 \\  
  9  & 22:47:50.43 & 63:56:46.6 &   7.86  &   1.32  & 0.620  &  3.910  & 4891 & 1.38  & 0.53 & 10$^*$  & 3.76 & 1.453 \\  
 10  & 22:47:45.65 & 63:56:27.1 & -23.59  & -18.15  & 0.560  &  1.377  & 3060 & 0.49  & 0.48 & 10$^*$  & 1.32 & 1.313 \\  
 11  & 22:47:49.40 & 63:56:52.1 &   1.10  &   6.84  & 0.470  &  1.066  & 2768 & 0.38  & 0.41 & 18  & 0.45 & 0.483 \\  
 12  & 22:47:51.22 & 63:56:38.4 &  13.08  &  -6.84  & 0.400  &  0.400  & 1510 & 0.14  & 0.35 & 57  & 0.04 & 0.103 \\  
 13  & 22:47:50.02 & 63:56:45.0 &   5.22  &  -0.29  & 0.370  &  0.370  & 1510 & 0.13  & 0.32 & 15  & 0.19 & 0.473 \\  
 14  & 22:47:51.19 & 63:57:00.3 &  12.86  &  14.99  & 0.360  &  0.517  & 2039 & 0.18  & 0.31 & 27  & 0.13 & 0.223 \\
 15  & 22:47:49.87 & 63:56:45.8 &   4.19  &   0.51  & 0.350  &  0.350  & 1510 & 0.12  & 0.30 & 20  & 0.12 & 0.303 \\
 \hline
   \multicolumn{10}{c}{ISOSS23053} \\
    \hline
  1  & 23:05:21.62 & 59:53:43.2 &  -6.41  &  -9.46  &18.380  & 62.675  & 8063 & 39.47 &15.80 & 186 & 3.29 & 1.32 \\  
  2  & 23:05:23.54 & 59:53:54.7 &   8.05  &   2.09  & 9.820  & 35.389  & 8018 & 22.29 & 8.46 & 139 & 2.51 & 0.95 \\  
  3  & 23:05:23.15 & 59:53:54.9 &   5.14  &   2.24  & 7.550  & 30.091  & 7324 & 18.95 & 6.50 & 67  & 4.61 & 1.58 \\  
  4  & 23:05:19.99 & 59:53:55.2 & -18.63  &   2.61  & 5.880  & 11.760  & 4916 &  7.41 & 5.07 & 33  & 3.95 & 2.70 \\  
  5  & 23:05:21.84 & 59:53:46.1 &  -4.77  &  -6.48  & 4.920  & 18.082  & 7055 & 11.39 & 4.24 & 112 & 1.61 & 0.60 \\  
  6  & 23:05:23.79 & 59:54:01.0 &   9.91  &   8.42  & 4.570  & 13.019  & 5793 &  8.20 & 3.94 & 88  & 1.50 & 0.72 \\  
  7  & 23:05:24.57 & 59:53:59.4 &  15.79  &   6.78  & 2.320  &  3.532  & 3349 &  2.22 & 2.00 & 54  & 0.69 & 0.62 \\  
  8  & 23:05:20.34 & 59:53:43.2 & -16.02  &  -9.39  & 2.290  &  2.937  & 2993 &  1.85 & 1.97 & 38  & 0.84 & 0.90 \\  
  9  & 23:05:22.41 & 59:53:46.2 &  -0.45  &  -6.41  & 1.880  &  2.519  & 2970 &  1.59 & 1.62 & 92  & 0.28 & 0.28 \\  
 10  & 23:05:20.74 & 59:53:52.1 & -13.04  &  -0.52  & 1.730  &  4.757  & 4264 &  3.00 & 1.49 & 30  & 1.80 & 0.90 \\  
 11  & 23:05:21.38 & 59:53:53.0 &  -8.20  &   0.37  & 1.500  &  1.500  & 1737 &  0.95 & 1.29 & 26  & 0.67 & 0.91 \\  
 12  & 23:05:19.82 & 59:54:02.6 & -19.97  &   9.98  & 1.460  &  1.887  & 2681 &  1.19 & 1.26 & 39  & 0.53 & 0.56 \\  
 13  & 23:05:23.78 & 59:53:51.7 &   9.83  &  -0.97  & 1.390  &  2.558  & 3262 &  1.61 & 1.20 & 61  & 0.44 & 0.32 \\  
 14  & 23:05:24.42 & 59:54:01.7 &  14.68  &   9.09  & 1.290  &  3.514  & 3808 &  2.21 & 1.11 & 116 & 0.30 & 0.15 \\
  \hline                                                                                              
  \hline
  \end{tabular}
  Notes: $^*$ The nominally derived temperature
  was either slightly below 10\,K or could not be well fitted (\#10). We set a 10\,K lower floor here.
  \end{center}
\end{table*}

\section{Results}
\label{results}

\subsection{Fragmentation and continuum emission}
\label{continuum}

To investigate the fragmentation properties of the large-scale gas
clumps visible in Figure \ref{overview}, we show the 1.3\,mm dust
continuum emission of the NOEMA data at a spatial resolution of
$\sim$0.8$''$ (Table \ref{lines}) in Figure \ref{cont}. Both regions
clearly fragment into numerous cores. Because we have no short spacing
information for the continuum emission, these cores appear to be separated
from each other. As we go on to see in the spectral line data discussed
below (Sect. \ref{kinematics}), these cores are nevertheless connected by
filamentary gas structures. Furthermore, we find that all cores are
associated with 70\,$\mu$m emission, hence, none of them are genuinely
starless any longer.

\begin{figure*}[htb]
\includegraphics[width=0.49\textwidth]{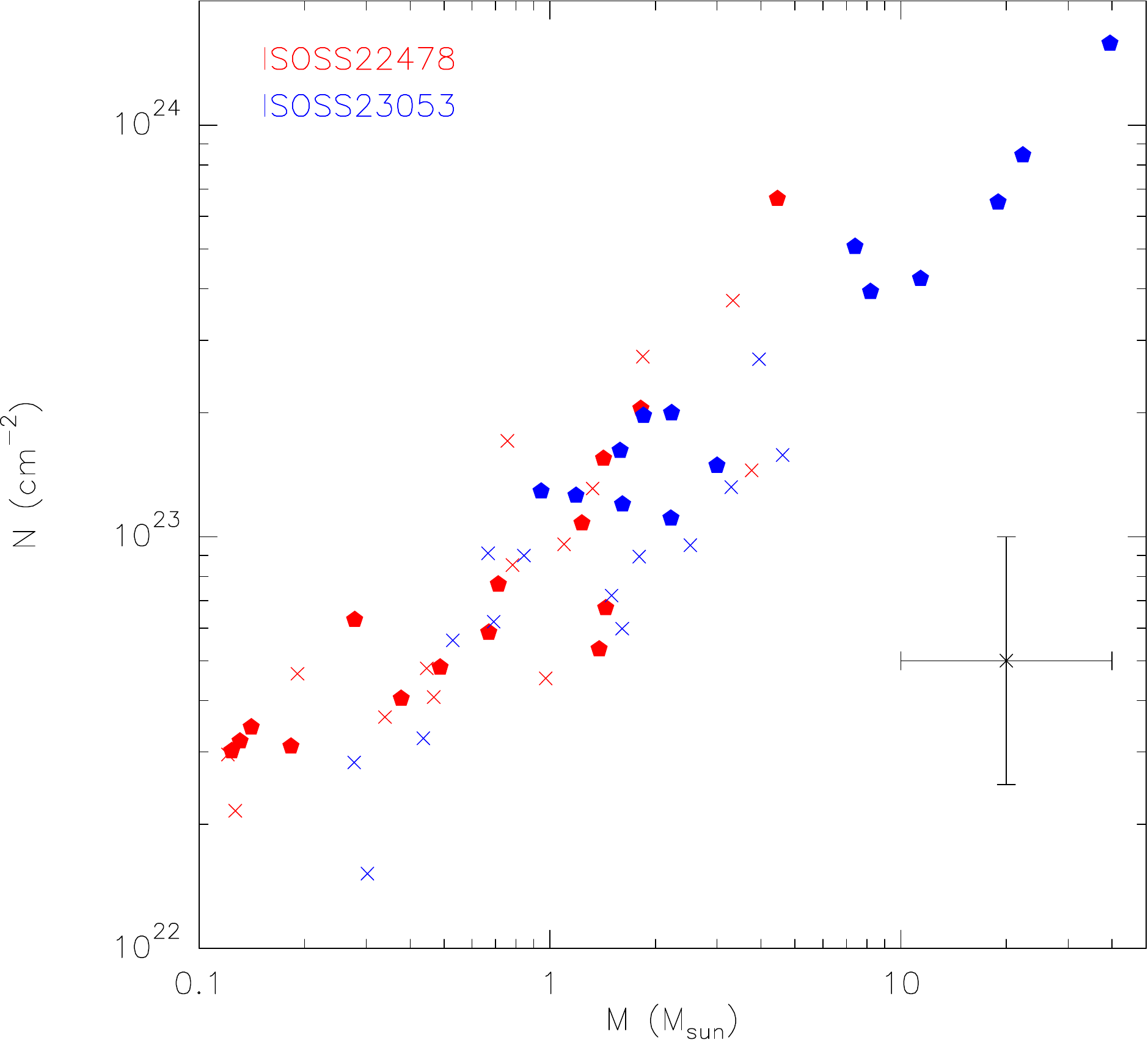}
\includegraphics[width=0.49\textwidth]{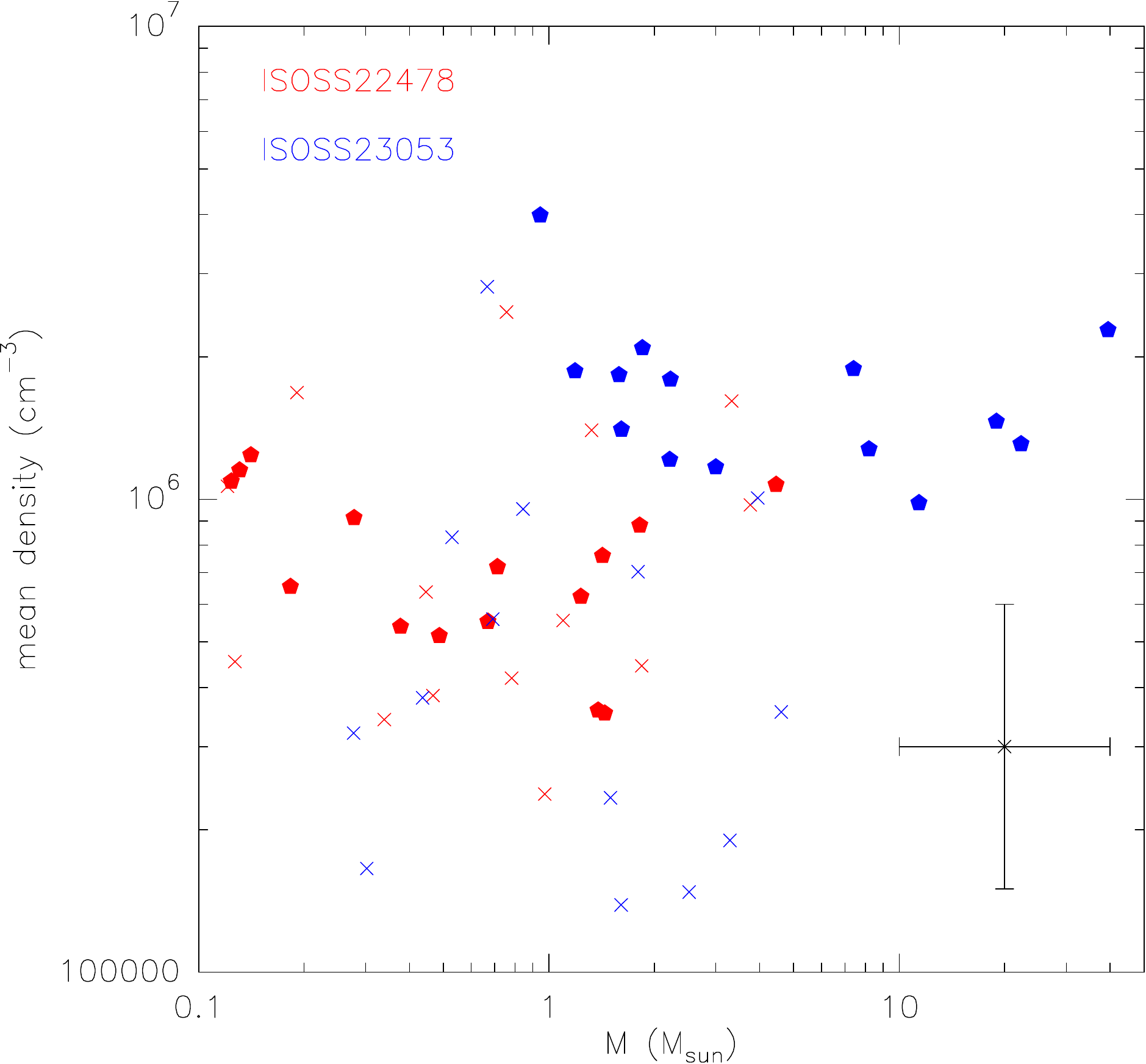}\\
\includegraphics[width=0.49\textwidth]{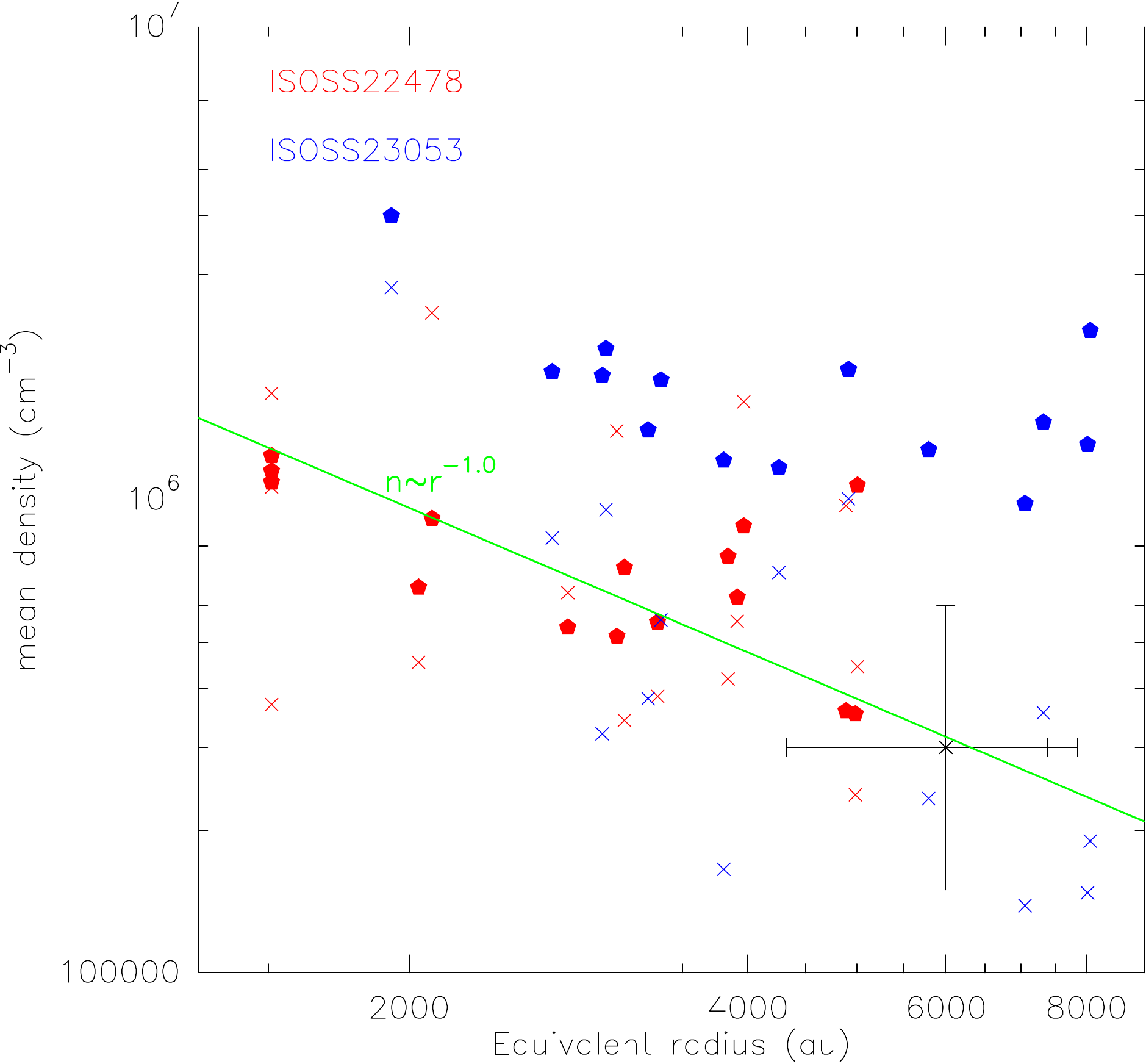}
\includegraphics[width=0.49\textwidth]{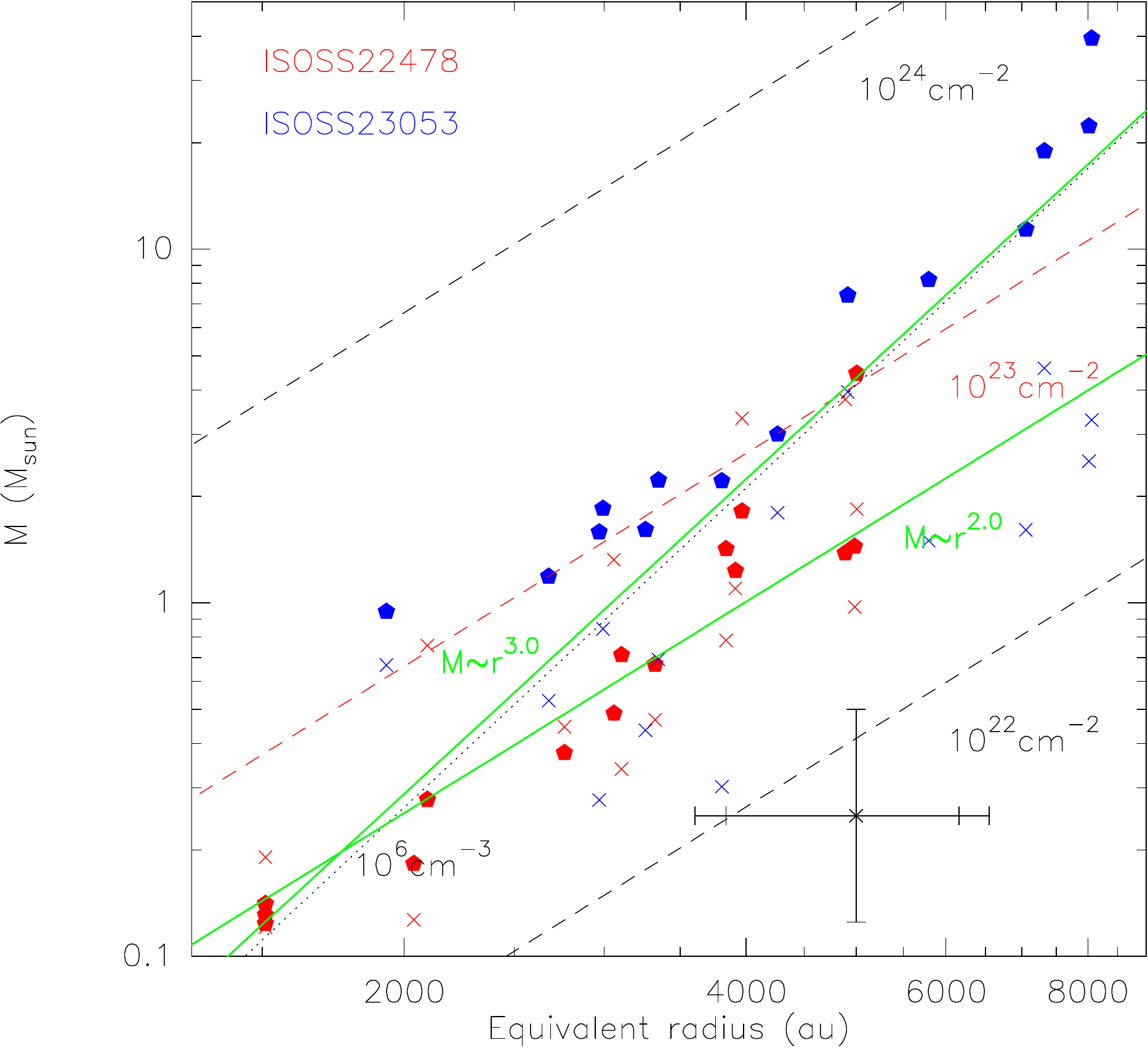}
\caption{Relations of core parameters. The top panels show the peak
  column density and mean density plotted against the mass while the
  bottom panels present the mean density and mass plotted against the
  equivalent radius. The red and blue colors always show the data for
  ISOSS22478 and ISOSS23053, respectively. The full pentagons show the
  results estimated assuming constant temperatures ($T_{\rm 20K}$)
  whereas the crosses present the data assuming the temperatures
  derived from the H$_2$CO line data ($T_{\rm H2CO}$).  The
  bottom-left panel presents as green line a fit to the data following
  the $T_{\rm H2CO}$ approach. The bottom-right panel also presents
  fits to the data (green continuous lines) and lines of constant mean
  column density and constant mean density (dashed and dotted lines,
  respectively) as labeled in the plot. Respective error-bars in each
  panel correspond to a factor of 2 uncertainty for masses, column
  densities, and mean densities, as well as 1\,kpc distance uncertainty
  for the equivalent radii (resulting in the different error-bar end
  markers for the radii).}
\label{relations} 
\end{figure*}

Although the number of cores is not excessively large, to
systematically derive the flux densities, offsets and sizes of the
cores, we use the {\sc clumpfind} algorithm originally introduced by
\citet{williams1994}. As a low-intensity threshold, we used the
$5\sigma$ contours presented in Fig.~\ref{cont_num}. We identify 15
and 14 cores for ISOSS22478 and ISOSS23053, respectively. Peak flux
densities $S_{\rm{peak}}$, integrated fluxes $S$, R.A.~ and Dec.,
offset positions ($\Delta x, \Delta y$) and equivalent core radii are
presented in Table \ref{cont_par}. The equivalent core radii are
calculated from the measured core area assuming a spherical
distribution. No deconvolution from the synthesized beam was applied.

Core masses and peak column densities can be estimated from the
integrated fluxes, $S,$ and the peak flux densities, $S_{\rm{peak}}$,
assuming optically thin dust continuum emission. Following
\citet{hildebrand1983} and \citet{schuller2009}, we use a gas-to-dust
mass ratio of 150 \citep{draine2011} and a dust mass absorption
coefficient $\kappa=0.9$\,cm$^2$g$^{-1}$ (\citealt{ossenkopf1994} at
densities of $10^6$\,cm$^{-3}$ with thin ice mantles).

The last bits of information needed is the temperature of the gas and
dust. For both regions several temperature estimates
exist. \citet{ragan2012b} used Herschel far-infrared data and fitted
the spectral energy distributions. This resulted in average dust
temperatures for ISOSS22478 and ISOSS23053 of $\sim$21 and
$\sim$22\,K, respectively. Both regions were also imaged with the Very
Large Array in NH$_3$ emission, and average temperatures from these
data are 13 and 18\,K for the two regions, respectively
\citep{bihr2015}. Within this new NOEMA dataset we also have the
H$_2$CO lines around 218\,GHz that can be used as thermometer as well
\citep{mangum1993}. Details about the temperature derivation and
structure are presented in Sect. \ref{kinematics}. Relevant for
our mass and column density analysis here is that for ISOSS22478 the
temperatures range approximately from values around 10\,K to roughly
50\,K (Table \ref{cont_par}). For ISOSS23053, the range is broader,
from around 10\,K to more than 100\,K. 
In Sect. \ref{kinematics}, we  discuss in more detail that a lot of the high temperature gas also
appears to be associated with shocks.

The above three temperature estimates indeed show some variance.
Since our mass and column density estimates are based on dust
continuum observations of dense gas cores, the Herschel dust, and the
NH$_3$ estimates may give good constraints. However, the new NOEMA
H$_2$CO temperature data have a higher spatial resolution and are also
able to trace a higher dynamic range in temperature than the dust and
NH$_3$(1,1)/(2,2) data. As discussed in Sect.~\ref{kinematics}, the
H$_2$CO temperatures appear to be also affected by the shock
structure. While the dust and NH$_3$ data may underestimate the
temperatures toward the main cores, the H$_2$CO data may overestimate
temperatures toward lower-mass cores in the regions that are close to
shock enhancements of the temperatures (e.g., core \#12 in ISOSS22478
or core \#14 in ISOSS23053, Table \ref{cont_par}). Hence, it is not a
priori clear which of the temperature estimates are better for
determining the masses and column densities. Therefore, in the
following, we derive our mass and column densities estimates with two
assumptions. Based on the dust and NH$_3$ temperature measurements, we
 use uniformly 20\,K for all mass and column density estimates in the first approach. In
the second, we use the temperatures derived from the H$_2$CO
data at each core position for the mass and column density
estimates. We refer to the two approaches as $T_{\rm 20K}$ and $T_{\rm
  H2CO}$.  While the interferometrically filtered out emission in the
continuum data implies that the measured fluxes are lower limits,
considering the additional uncertainties in the temperature estimate,
distances, dust absorption coefficient $\kappa$ as well as the
gas-to-dust mass ratio, we estimate the masses and column densities to
be accurate within a factor of 2-4 (see also the CORE study of the original
sample, \citealt{beuther2018b}).

The derived core masses ($M_{\rm 20K}$ and $M_{\rm T_{\rm H2CO}}$),
peak column densities ( $N_{\rm peak, 20K}$ and $N_{\rm peak, T_{\rm
    H2CO}}$), and H$_2$CO temperatures ($T_{\rm H2CO}$) for the two
approaches outlined above are presented in Table \ref{cont_par} for
all 29 cores (15 and 14 cores for ISOSS22478 and ISOSS23053,
respectively). The range of core masses for ISOSS22478 is between
$\sim$0.05 and $\sim$4.5\,M$_{\odot}$ whereas core masses for
ISOSS23053 range between $\sim$0.3 and 39.5\,M$_{\odot}$, depending
also on the temperature assumptions. While the difference in ranges
may partially be attributed to the slightly larger distance of
ISOSS23053, this region is also far more massive in general (Sect.
\ref{targets}). Hence, it is not too surprising that more massive
cores are detected in that region. The column densities for both
regions range between a few times $10^{22}$ to
$\sim$10$^{24}$\,cm$^{-2}$. With the given assumptions at 20\,K the
$3\sigma$ mass and column densities for ISOSS22478 are
0.06\,M$_{\odot}$ and $1.5\times 10^{22}$\,cm$^{-2}$,
respectively. The corresponding $3\sigma$ values again at 20\,K for
ISOSS23053 are 0.3\,M$_{\odot}$ and $4.1\times 10^{22}$\,cm$^{-2}$.

The total mass in all cores for ISOSS22478 and ISOSS23053 are 14.7 and
121.9\,M$_{\odot}$ assuming a temperature of 20\,K,
respectively. Comparing these values with the total mass reservoir
available in the regions (Sect. \ref{targets}), because there is no
short-spacing information for the continuum data, at most 10\%
and 20\% of the continuum flux can be recovered by the NOEMA
observations for ISOSS22478 and ISOSS23053, respectively.

Figure \ref{relations} presents several relations derived for the
cores of the two regions. All plots show the derived parameters with
both temperature assumptions. The top-left panel shows the column
density against the core mass, and we find a clear correlation between
the two parameters, independent of the temperature assumption. While
for the $T_{\rm H2CO}$ approach, the masses and column densities
between the two regions overlap strongly, in the $T_{\rm 20K}$
approach, the estimated masses and column densities are larger for
ISOSS23053. Nevertheless, also in the $T_{\rm 20K}$ approach, there is
clear overlap between both regions and the general trend for these two
quantities agrees well.

Employing the equivalent radius of the cores from the {\sc clumpfind}
analysis, we can also derive mean densities for all cores assuming
spherical symmetry. The top-right and bottom-left panels of
Fig.~\ref{relations} plot these mean densities versus the core mass
and the core size, respectively. While the scatter in the $T_{\rm
  20K}$ approach for the densities is smaller around almost constant
densities of $\sim 10^6$\,cm$^{-3}$, the bottom-left panel of
Fig.~\ref{relations} is indicative of a decreasing trend of densities
with increasing equivalent core radius in the $T_{\rm H2CO}$
approach. Fitting a power-law to the size-density relation in this
$T_{\rm H2CO}$ approach, we find a relation of $n\propto r^{-1.0\pm
  0.3}$. We note that in the classical 3rd Larson relation,
\citet{larson1981} found that for his sample of CO molecular clouds
the mean density is inversely related to the size ($n\propto
r^{-1.1}$).

\begin{table}
\caption{Linear minimum spanning tree analysis}
\begin{tabular}{lccccc}
\hline \hline
Source & \#cores& mean sep & min sep & max sep \\
       &            & (au)      & (au)     & (au) \\
\hline\hline
ISOSS22478 & 15 & 22110 & 3890  & 73579 \\
ISOSS23053 & 14 & 24630 & 11046 & 44389 \\
\hline
\hline
\end{tabular}
\label{min_span_tree}
\end{table}

The bottom-right plot of Fig.~\ref{relations} shows the core masses
against the equivalent radius, that is, the mass-size relation.  While we
see a good correlation between these two quantities, the slopes for
the relation derived under the two temperature assumptions varies
significantly. The mass-size relation in the $T_{\rm H2CO}$ approach
follows $M\propto r^{2.0\pm 0.3}$, close to the lines of constant
column densities as would be expected from Larson's relations (e.g.,
\citealt{heyer2009,lombardi2010,ballesteros2019,ballesteros2020}). In
contrast to this, the mass-size relation in the $T_{\rm 20K}$ approach
follows a steeper relation of $M\propto r^{3.0\pm 0.2}$, which is
close to the line of constant density at $10^6$\,cm$^{-3}$.  If the
two regions are separated in the $T_{\rm 20K}$ approach, the mass-size
relations for ISOSS22478 and ISOSS23053 have power-law dependences of
$r^{2.4\pm 0.2}$ and $r^{2.6\pm 0.2}$, respectively. We return to
this point in Sect. \ref{fragmentation}.

\begin{figure}[htb]
\includegraphics[width=0.49\textwidth]{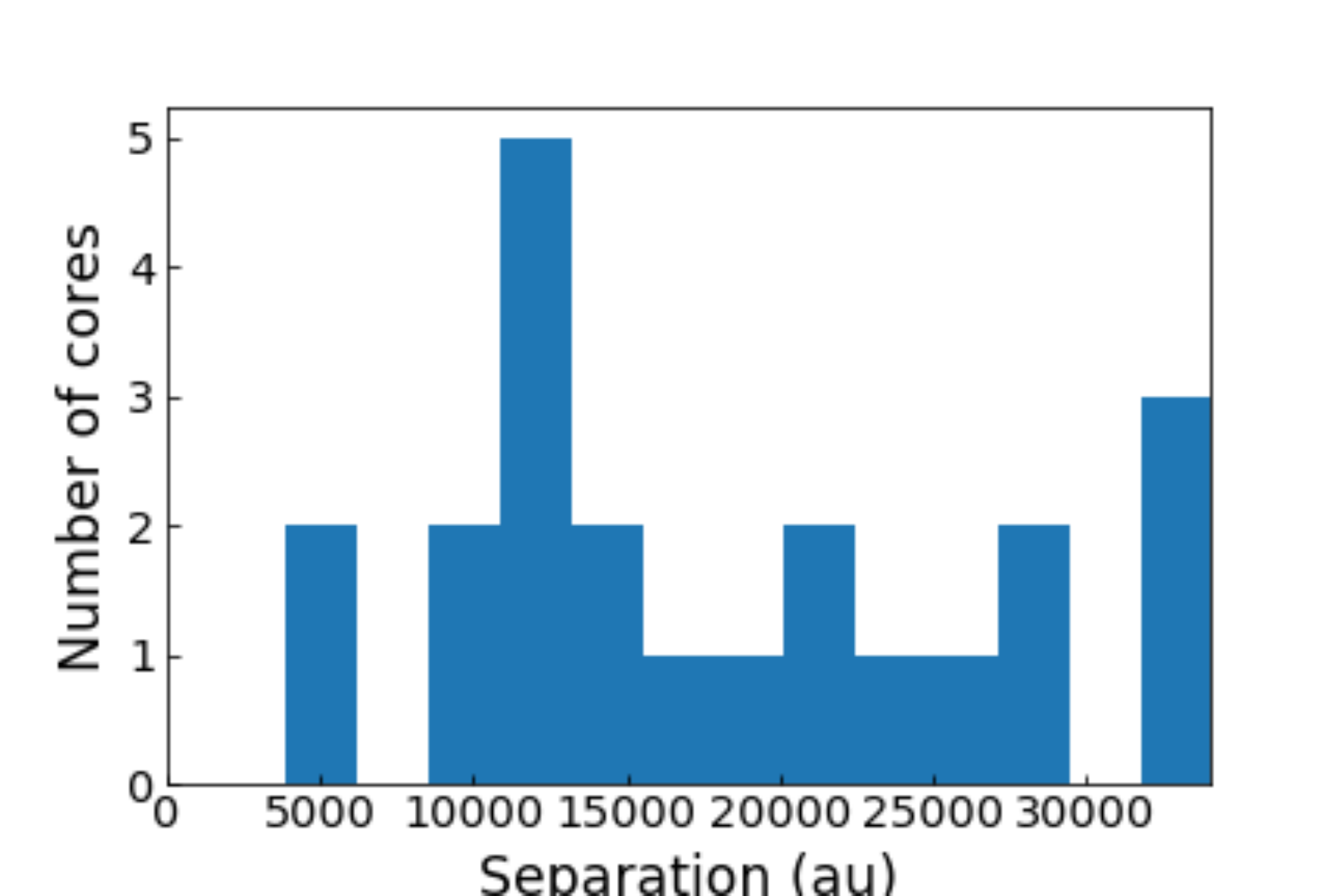}
\caption{Nearest-neighbor separation histogram from minimum spanning
  tree analysis for both regions combined. The spatial resolution
  limits are $\sim$2700 and $\sim$3500\,au for ISOSS22478 and
  ISOSS23053, respectively.}
\label{global_sep} 
\end{figure}

To investigate the projected separations between cores, we employed a
minimum spanning tree analysis using the \verb+astroML+ software
package \citep{vanderplas2012} that determines the shortest possible
distances connecting each of the cores. The derived nearest-neighbor
separations are shown in Fig.~\ref{global_sep} for both regions
combined, and the basic results are listed in Table
\ref{min_span_tree}. While projection causes the observed
nearest-neighbor separations to be at their lower limits, as compared
to their real separation in three dimensions, the sensitivity, and
spatial resolution limits that the observation are subject to,
implying that the projected nearest-neighbor separations are, rather,
likely to be the upper limits. In reality, having more cores should
qualitatively reduce the real projected nearest-neighbor
separations. Nevertheless, with a linear spatial resolution of
approximately 2700 and 3500\,au, it is interesting to note that barely
any separations are found at our resolution limit. This is different
in comparison to the original CORE sample of more evolved HMPOs where
the nearest-neighbor histogram shows a strong peak near the resolution
limit. While also for the younger ISOSS sources, we would expect that
at higher spatial resolution a few of the cores are likely to split
into multiple systems and, hence, some smaller separations would
occur, the current data nevertheless show that we are resolving the
large-scale fragmentation of the parental gas clumps well. A decrease
in spatial separations during the evolution of a star-forming cluster
is expected during the collapse of the gas clump. During the collapse,
the densities increase and in the classical Jeans picture, the Jeans
length and the corresponding separations decrease. In that scenario,
the generally larger separations of the younger ISOSS sources studied
here can naturally be explained.

It is interesting to note that in a study of the integral shape
filament in Orion, \citet{kainulainen2017} found that significant
grouping of dense cores is found below scales of 17000\,au (see also
\citealt{roman-zuniga2019}), similar to what we find in
Fig.~\ref{global_sep}. \citet{kainulainen2017} infer a second increase
in source grouping below separations of 6000\,au, which is not visible
in our data. However, smaller-scale grouping (below our resolution
limit here) was found in the original CORE data at approximately
twice the spatial resolution with a peak at roughly 2000\,AU
\citep{beuther2018b}. This missing smaller-scale grouping in the two
regions studied here may potentially be attributed to insufficient
spatial resolution. We come back to that in Sect.
\ref{fragmentation}.

\begin{figure*}[htb]
\includegraphics[width=0.99\textwidth]{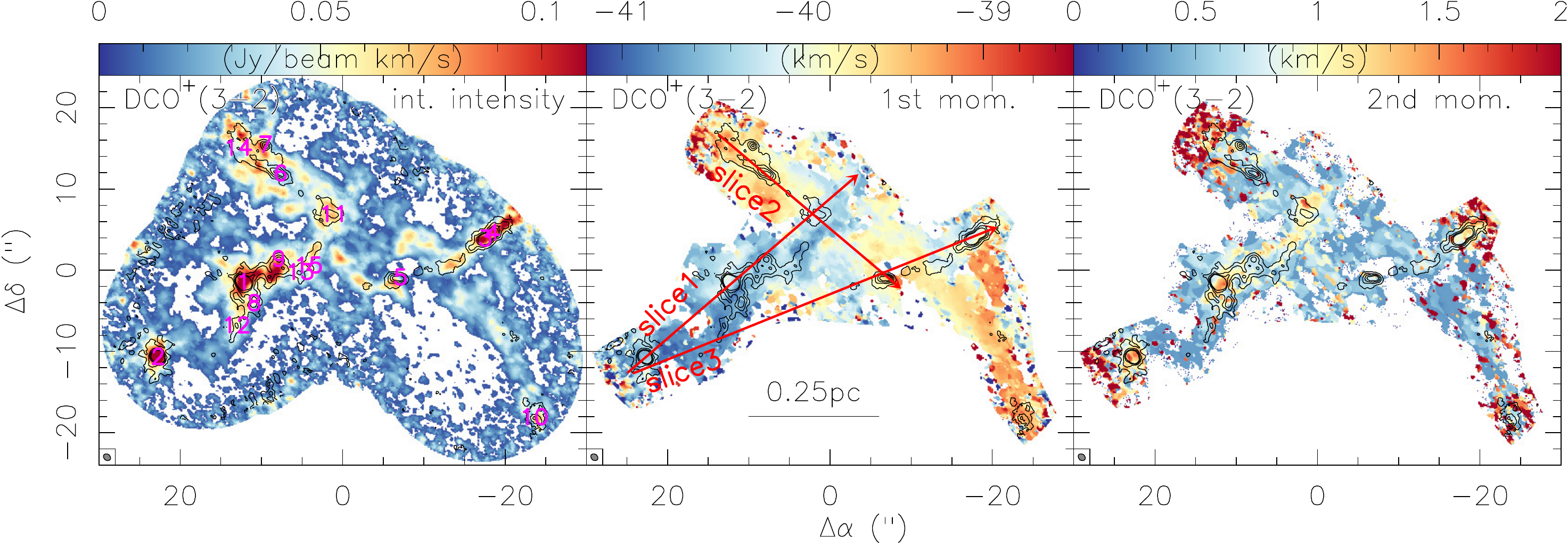}
\caption{NOEMA+30\,m DCO$^+$(3--2) data towards ISOSS22478. The three
  panels show, in color scale the integrated intensity, the 1st and 2nd
  moment maps (intensity-weighted peak velocities and velocity
  dispersions), respectively. These maps were produced by clipping all
  data below an approximate $3\sigma$ threshold of
  18\,mJy\,beam$^{-1}$. The contours show the 1.3\,mm continuum
  emission in $3\sigma$ steps ($1\sigma \sim 0.057$\,mJy\,beam$^{-1})$
  from 3 to $15\sigma$. The beam is shown in the bottom-left of all
  panels, and a linear scale bar is shown in the middle panels. The
  red arrows in the middle panel show the three position velocity
  slices presented in Fig.~\ref{22478_slices}. To reduce noise
  signatures at the edges of the mosaic, for the 1st and 2nd moment
  maps, we blanked the mosaic edges. The cores are labeled in the left
  panel.}
\label{22478_dco+} 
\end{figure*}

\subsection{Kinematics and temperatures of the gas}
\label{kinematics}

We consider what the kinematics of the gas tell us about the formation
processes within these two star-forming regions. To look at the
dynamics in more detail, in the following we focus on the
DCO$^+$(3--2) emission that traces the gas properties at the given
early evolutionary stage extremely well (e.g.,
\citealt{gerner2015}). Figures \ref{22478_dco+} and \ref{23053_dco+}
present an overview of the DCO$^+$ data. The left panels show the
integrated intensities and the middle and right panels present the 1st
and 2nd moment maps (intensity-weighted peak velocities and velocity
dispersions), respectively. These are always the merged datasets
combining the NOEMA ACD configurations with the IRAM 30\,m data
(Sect. \ref{data}).

\subsubsection{ISOSS22478}

\begin{figure}[htb]
\includegraphics[width=0.49\textwidth]{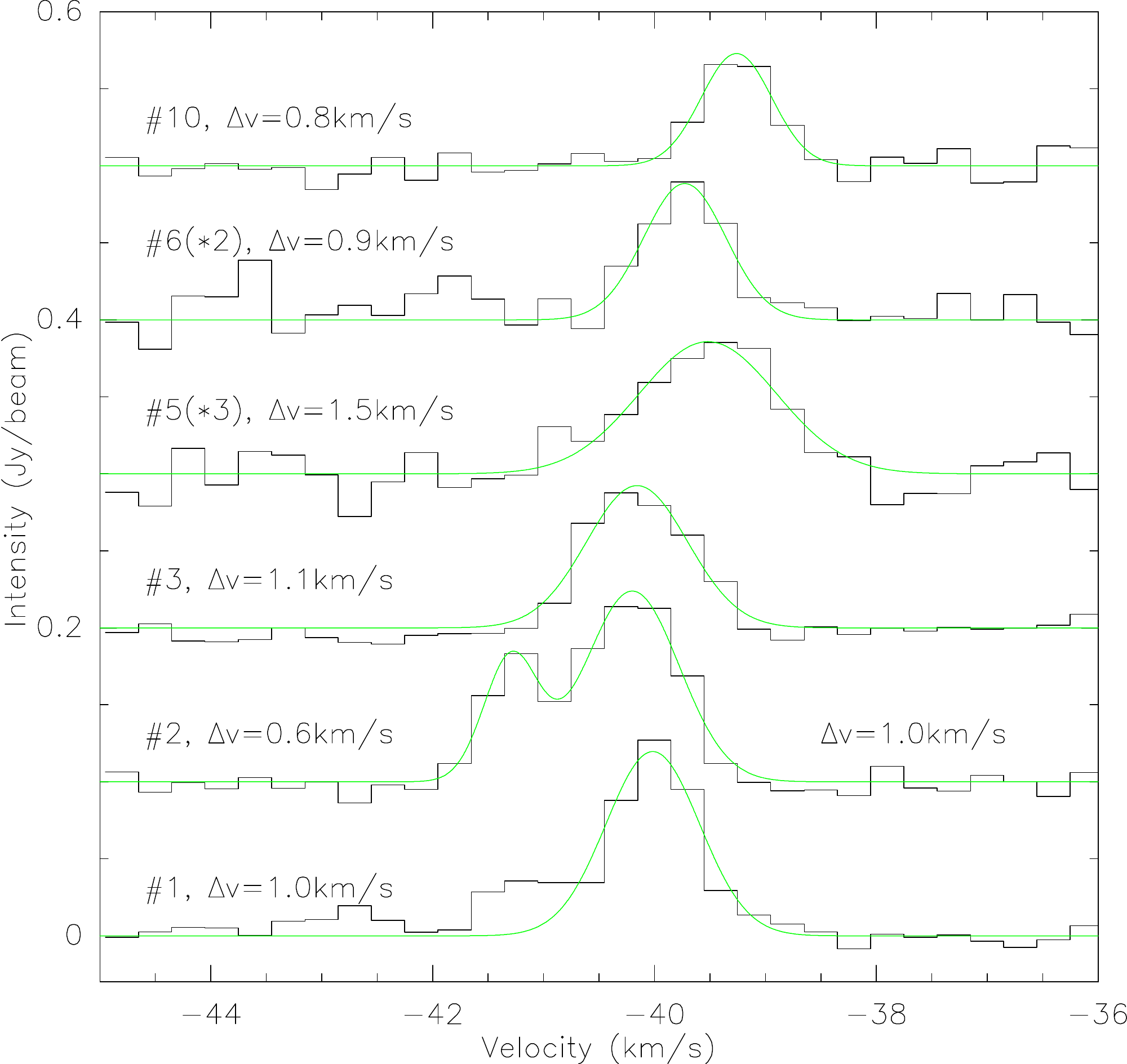}
\caption{DCO$^+$(3--2) example spectra toward the labeled core
  positions in ISOSS22478 (Fig.~\ref{cont_num}). The spectra are
  shifted on the Y-axis for better presentation. The spectra of core
  \#5 and core \#6 are multiplied by 3 and 2 for clarity. The green
  lines show Gaussian fits, the FWHM linewidth
  $\Delta \varv$ are presented for all spectra. For core \#2, two
  components were fitted.}
\label{spec_22478} 
\end{figure}

For ISOSS22478, the integrated emission clearly shows extended
filamentary gas emission structures that connect the 1.3\,mm continuum
cores. The 1st moment map in the middle panel of Fig.~\ref{22478_dco+}
reveals a complex velocity structure with slightly different velocity
components for the different sub-regions. We return to this
point below. The velocity dispersion shown in the right panel shows on
average low values largely around or even below 1\,km\,s$^{-1}$ (see
also the cases of G29.96 and G35.20 for similarly narrow lines in
deuterated ammonia, NH$_2$D, \citealt{pillai2011}). For comparison,
Fig. \ref{spec_22478} presents individual example spectra toward six
cores. While the spectrum toward core \#2 is better fitted with two
components, some other spectra are still consistent with
single-component fits. The full width half maximum (FWHM) $\Delta \varv$
values are on average small and vary roughly between 0.6 and
1.5\,km\,s$^{-1}$ (thermal line width of DCO$^+$ at 20\,K $\sim
0.17$\,km\,s$^{-1}$.)

\begin{figure*}[htb]
\includegraphics[width=0.99\textwidth]{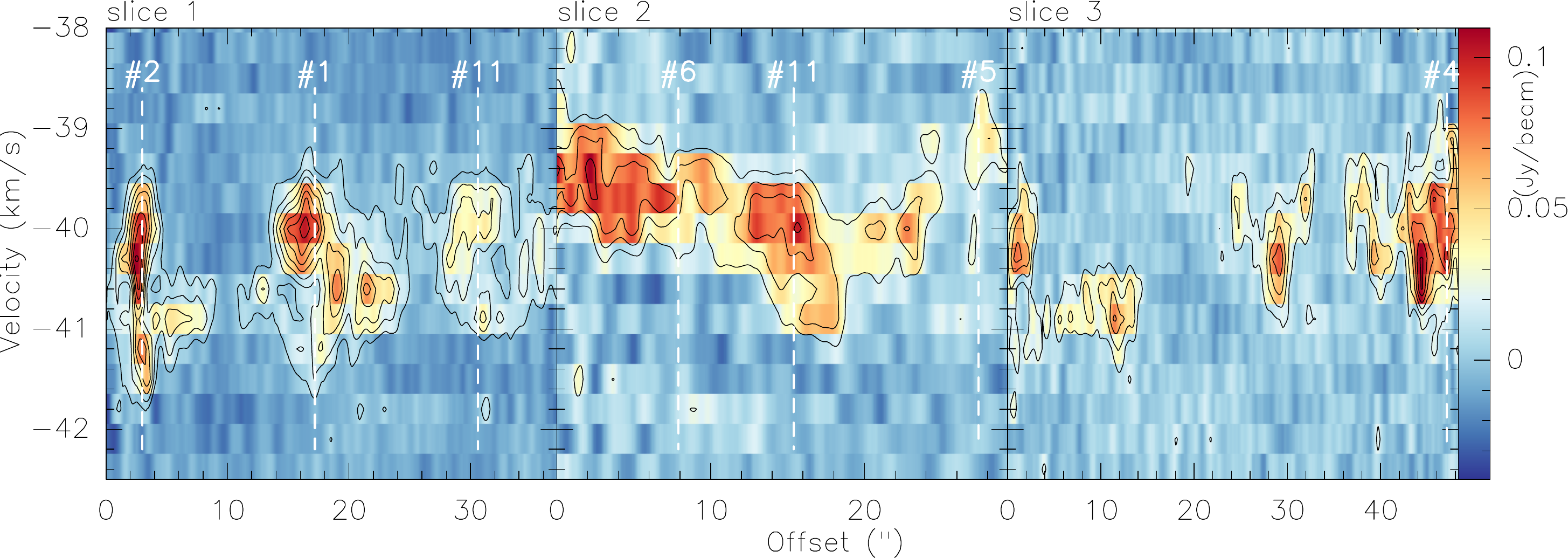}
\caption{Position-velocity diagrams in DCO$^+(3-2)$ for source
  ISOSS22478 along the cuts shown in Fig.~\ref{22478_dco+}. The
  contours are in $3\sigma$ steps of 18\,mJy\,beam$^{-1}$. The
  vertical dashed lines mark the positions of a few selected cores.}
\label{22478_slices} 
\end{figure*} 

A different way to look at the kinematics of the gas is to consider them as
position-velocity slices along specific cuts within a region. Here, we
concentrate on three slices marked in the middle panel of
Fig.~\ref{22478_dco+}. Slice 1, connecting mainly cores \#2 to \#1 and
\#11, exhibits interesting multiple velocity components towards most
of the core positions along the slice. Similarly, slice 3 also shows
several velocity components at individual positions along the cut. In
contrast to these two slices, the almost perpendicular cut, connecting
mainly cores \#6, \#11, and \#5 shows a different velocity gradient,
with initially decreasing and then increasing velocities.  Multiple
velocity components at core positions have already been found in a few
other regions (e.g., G035.39, G35.2, IRDC\,18223, Orion,
\citealt{henshaw2014,sanchez2014,beuther2015b,hacar2017}). While such
multiple velocity components can indicate the internal dynamics of the
cores, for example the presence of rotating structures (e.g., G35.2,
\citealt{sanchez2014}), these signatures have often been interpreted
in the framework of fibers or interacting gas sheets. We
return to this in Sect. \ref{dynamics}.

\begin{figure}[htb]
\includegraphics[width=0.49\textwidth]{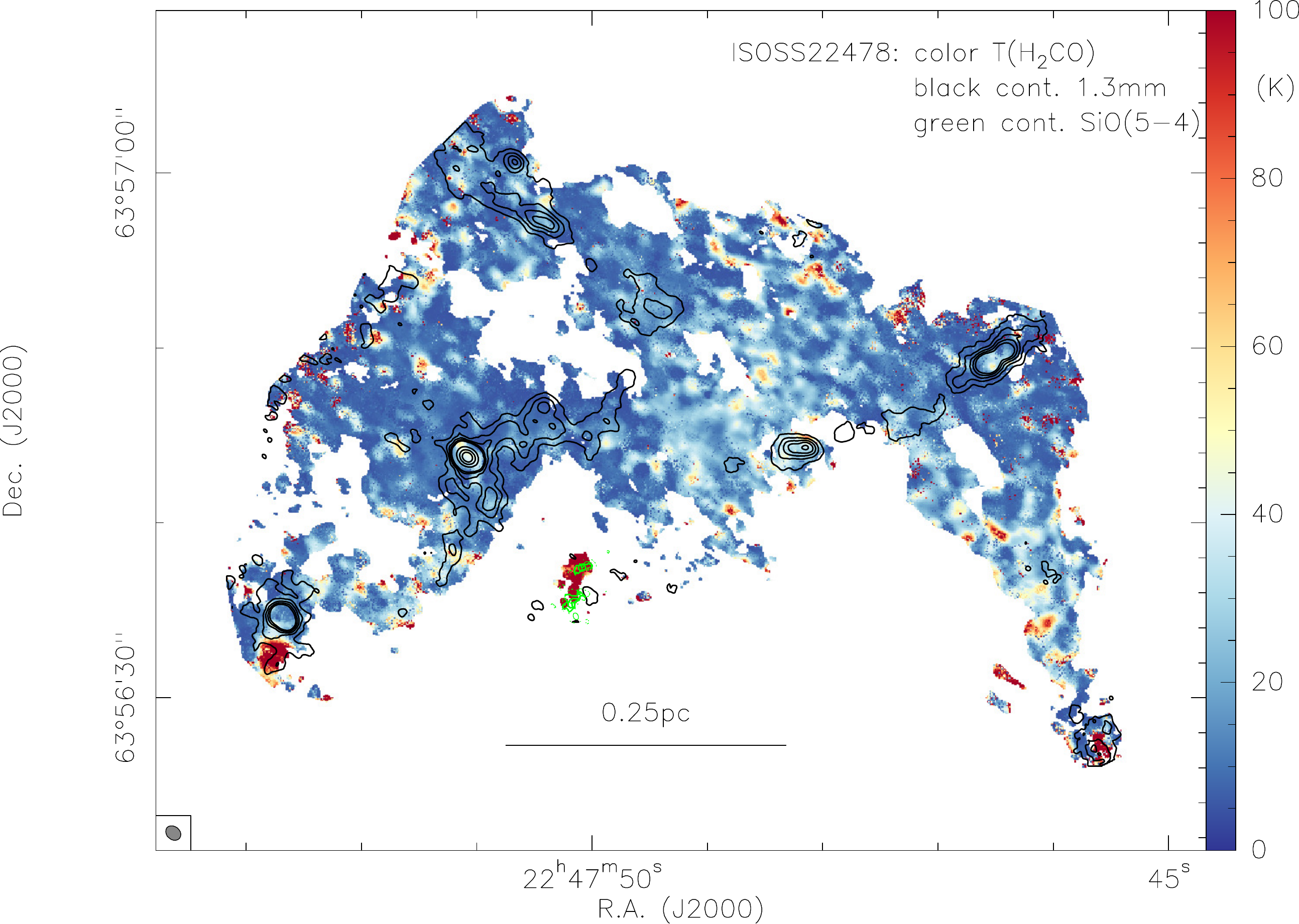}
\caption{Temperature map for ISOSS22478 derived from the H$_2$CO line
  emission shown in color scale. The edges of the map are blanked
  because of the lower signal-to-noise ratio. The black contours
  present the 1.3\,mm continuum starting at the $3\sigma$ level
  ($1\sigma \sim 0.057$\,mJy\,beam$^{-1})$ up to 15$\sigma$ and then
  continue in 30$\sigma$ steps. The green contours show SiO emission
  integrated from -44 to -38\,km\,s$^{-1}$ clipping all data below an
  approximate $4\sigma$ threshold of 30\,mJy\,beam$^{-1}$. The contour
  level is 0.2\,Jy\,beam$^{-1}$\,km\,s$^{-1}$. The beam is shown in
  the bottom-left, and a linear scale bar is shown as well.}
\label{t_22478} 
\end{figure} 

Another important parameter is the temperature of the gas in the
region. Figure \ref{t_22478} presents the temperature map we derived
from the H$_2$CO lines around 218\,GHz (Table \ref{lines}), which are
a well-characterized thermometer of the interstellar gas (e.g.,
\citealt{mangum1993,rodon2012,gieser2019}). To derive this temperature
map, we used the eXtended CASA Line Analysis Software Suite ({\sc
  xclass},
\citealt{moeller2017}\footnote{https://xclass.astro.uni-koeln.de})
tool within the {\sc casa} software package \citep{mcmullin2007}.  In
{\sc xclass} the spectra are fitted pixel by pixel taking also the
optical depth into account. The mean uncertainties for the
temperatures are on the order of 30\%. More details on the {\sc
  xclass} fitting are given in Gieser at al.~(subm.~to A\& A). In
addition to this, an in-depth analysis of the chemical and temperature
structure of cores within the the two regions will be presented in
Gieser et al.~(in prep.). As we can see in Fig.~\ref{t_22478}, the
temperatures in ISOSS22478 are homogeneously very low, typically below
20\,K, and only in the central part of the map do they rise above
20\,K over some area.  The temperature enhancement above 50\,K in the
small condensation in the south of the map (near the green contours in
Fig.~\ref{t_22478}) is considered to be real because it is the only
spot where also SiO(5--4) emission is detected. Hence, while we find
at one localized position a temperature enhancement that seems to be
associated with shocks, as suggested by the detection of SiO emission,
we do not identify strong temperature differences along the main
filamentary parts of the cloud. This  is set into contrast to the
SiO emission and temperature structure in ISOSS23053 below.

\begin{figure*}[htb]
\includegraphics[width=0.99\textwidth]{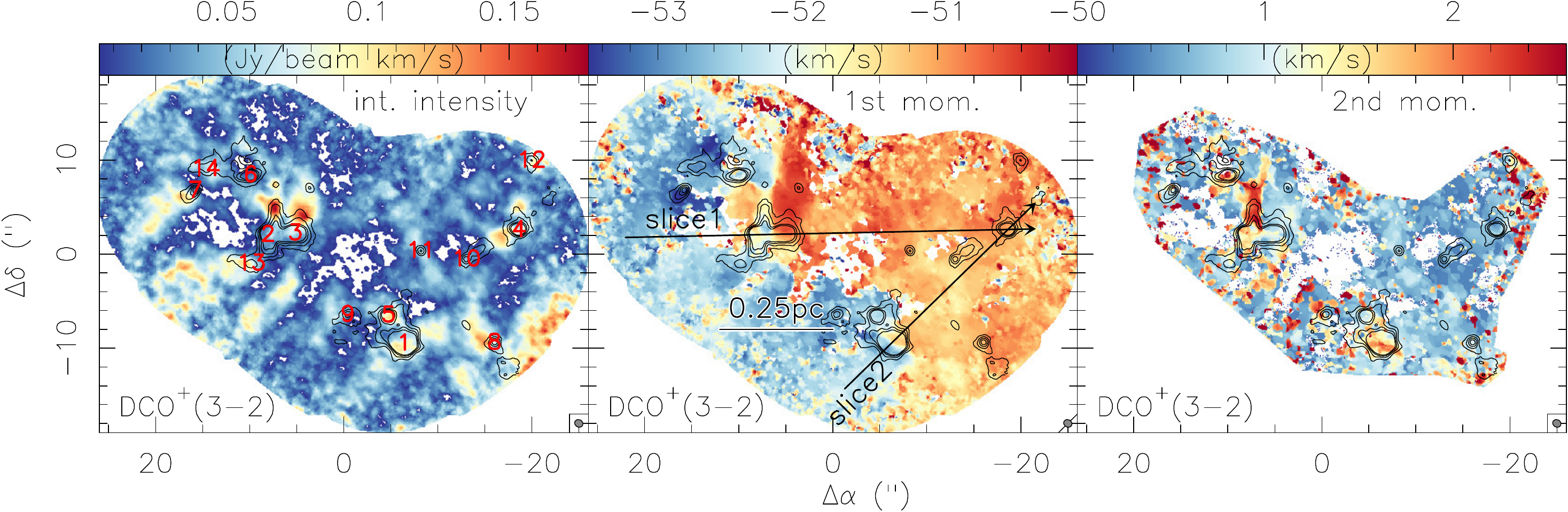}
\caption{NOEMA+30\,m DCO$^+$(3--2) data towards ISOSS23053. The three
  panels show in color scale the integrated intensity, the 1st and 2nd
  moment maps (intensity-weighted peak velocities and velocity
  dispersions), respectively. These maps were produced by clipping all
  data below an approximate $3\sigma$ threshold of
  15\,mJy\,beam$^{-1}$. The contours show the 1.3\,mm continuum
  emission in $3\sigma$ steps ($1\sigma \sim 0.16$\,mJy\,beam$^{-1})$
  from 3 to $15\sigma$. The beam is shown in the bottom-right of all
  panels, and a linear scale bar is shown in the middle panel. The
  arrows in the middle panel show the two position velocity slices
  presented in Fig.~\ref{23053_slices}. To reduce noise signatures at
  the edges of the mosaic, for the 2nd moment map, we blanked the
  mosaic edges. The cores are labeled in the left panel.}
\label{23053_dco+} 
\end{figure*} 

\subsubsection{ISOSS23053}

In comparison to ISOSS22478, the DCO$^+$ gas emission in ISOSS23053
looks very different (Fig.~\ref{23053_dco+}). Visually, a filamentary
structure is less apparent. In comparison to ISOSS22478, where the
DCO$^+$ and mm continuum emission spatially agree relatively well, for
ISOSS23053 many of the DCO$^+$ peaks are offset from the main mm
continuum peak positions. At the given spatial resolution, the
corresponding continuum peak brightness temperatures are all below
1\,K and the dust continuum emission is optically thin. Hence, dust
optical depth effects cannot explain these offsets. More important in
the context of the dynamical state of the region, the 1st moment map
in Fig.~\ref{23053_dco+} exhibits two very distinct velocity
components that merge in the middle of the region on scales below
0.1\,pc, right where we find most of the strong mm continuum
cores. These two velocity components were already identified in
previous NH$_3$ data, and \citet{bihr2015} suggested that they may be
the signature of two converging gas streams that collide and trigger a
new generation of star formation.

\begin{figure}[htb]
\includegraphics[width=0.49\textwidth]{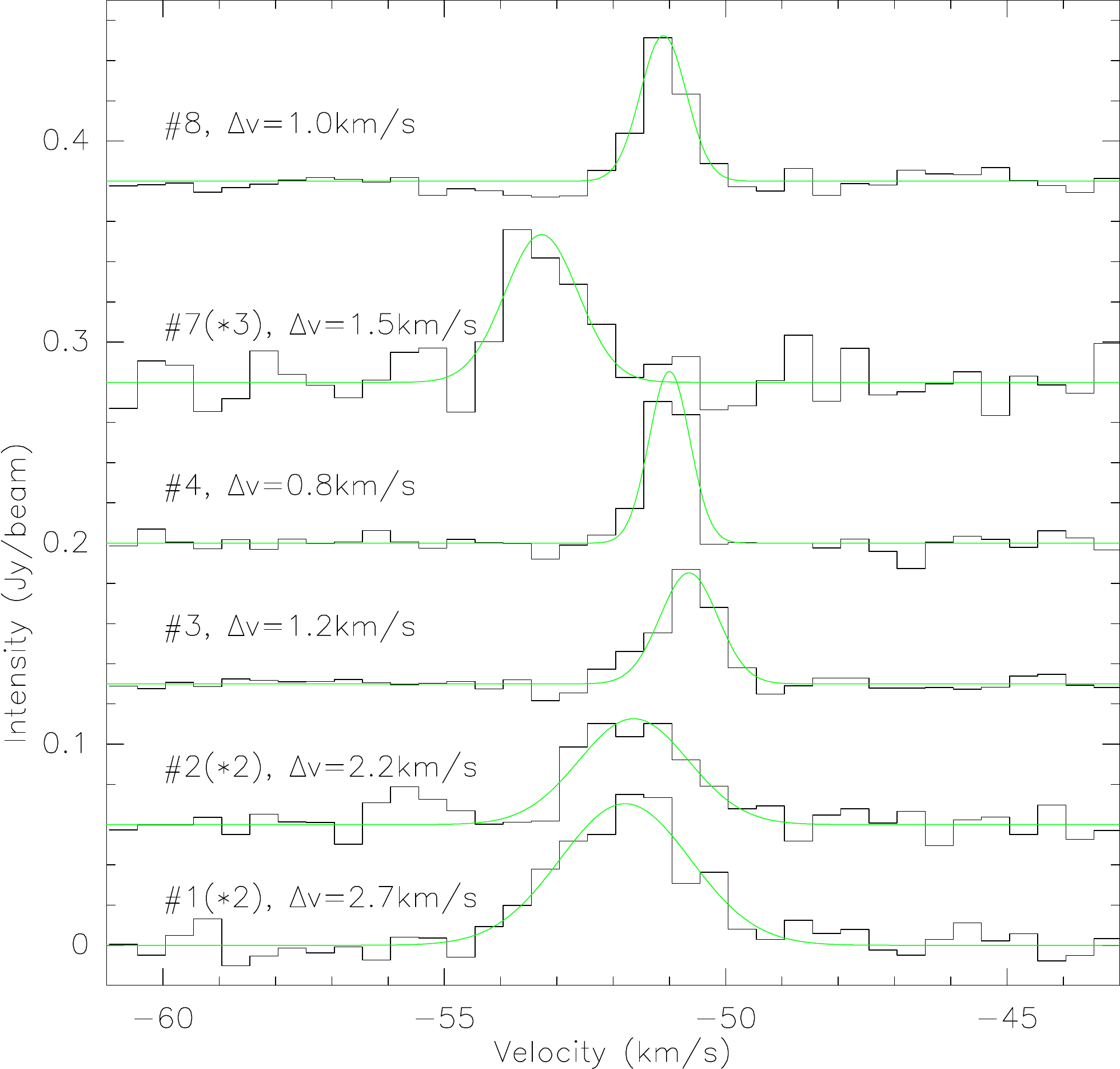}
\caption{DCO$^+$(3--2) example spectra toward the labeled core
  positions in ISOSS23053 (Fig.~\ref{cont_num}).  The spectra are
  shifted on the Y-axis for better presentation. The spectra of core
  \#1, \#2, and \#7 and are multiplied by 2, 2, and 3 for clarity. The
  green lines show Gaussian fits and the FWHM
  linewidth $\Delta \varv$ is presented for all spectral lines.}
\label{spec_23053} 
\end{figure}

In that context, it is interesting to investigate the velocity
dispersion. The 2nd moment map in Fig.~\ref{23053_dco+} reveals that
the velocity dispersion is largely low, at or below
1\,km\,s$^{-1}$. However, we see an increase in velocity dispersion
above 1\,km\,s$^{-1}$ towards the main cores, and even higher values a
bit to the north of the two main central cores (\#2 \& \#3). Comparing
these moment maps to a few selected spectra (Fig.~\ref{spec_23053}),
most of the spectra have rather narrow full width half maximum $\Delta
\varv$ values below 1.5\,km\,s$^{-1}$ with the exception of cores \#1
and \#2 that reveal $\Delta \varv$ above 2\,km\,s$^{-1}$. The spectra
also show clearly the two velocity components, for instance, core \#7 peaks
around -54\,km\,s$^{-1}$ whereas the emission peak of core \#8 is
rather at -51\,km\,s$^{-1}$.

\begin{figure*}[htb]
\includegraphics[width=0.99\textwidth]{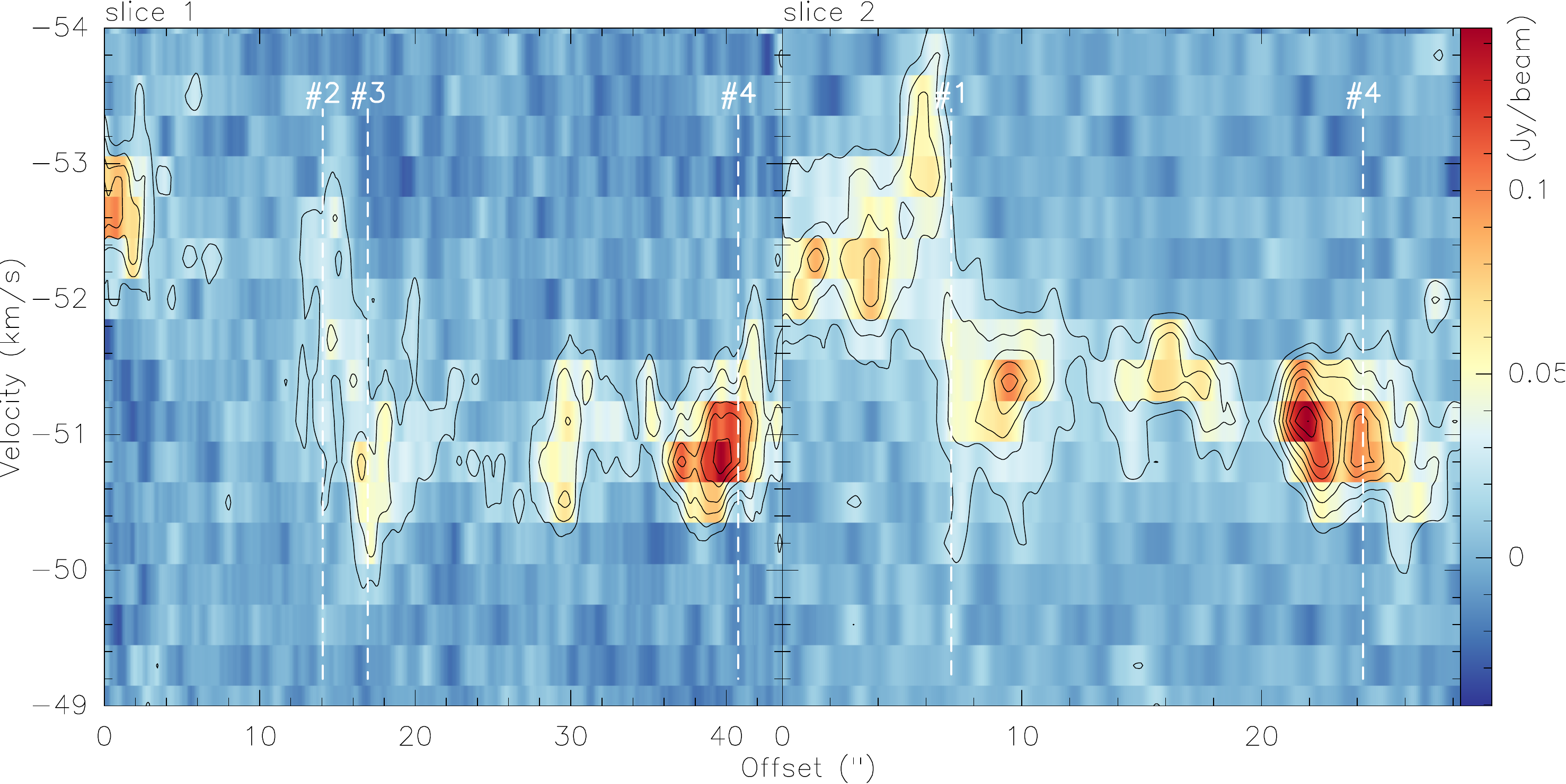}
\caption{Position-velocity diagrams in DCO$^+(3-2)$ for source
  ISOSS23053 along the cuts shown in Fig.~\ref{23053_dco+}. The
  contours are in $3\sigma$ steps of 18\,mJy\,beam$^{-1}$. The
  vertical dashed lines mark the positions of a few selected cores.}
\label{23053_slices} 
\end{figure*}

Regarding the two velocity components in ISOSS23053,
Fig.~\ref{23053_slices} shows two position-velocity slices with the
directions outlined in the middle panel of Fig.~\ref{23053_dco+}. Both
slices clearly show the two velocity components that connect on very
small spatial scales. While \citet{bihr2015} report this velocity step
as unresolved in their $\sim$4$''$ NH$_3$ observations and give a
lower limit to the gradient of $>30$\,km\,s$^{-1}$\,pc$^{-1}$, our
higher spatial resolution allows us to better quantify this
structure. For slices 1 and 2, we measure velocity gradients of
$\sim$48\,km\,s$^{-1}$\,pc$^{-1}$ and
$\sim$54\,km\,s$^{-1}$\,pc$^{-1}$, respectively. They can be even
larger in the SiO emission discussed in the following.

\begin{figure*}[htb]
\includegraphics[width=0.99\textwidth]{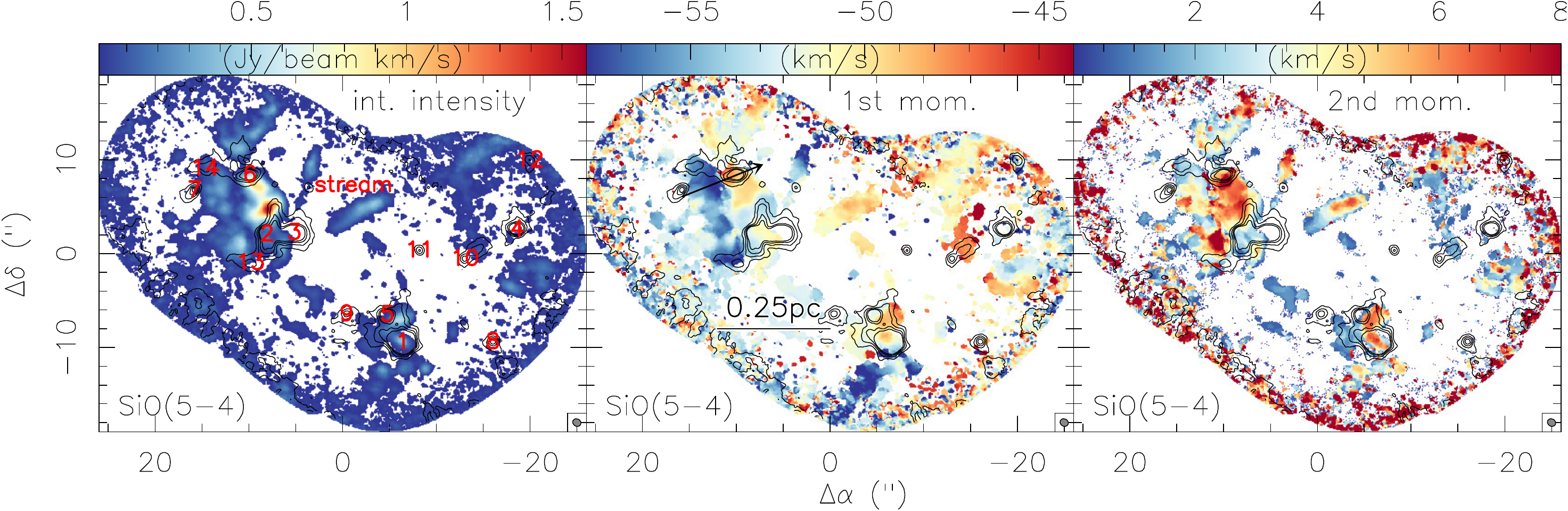}
\caption{NOEMA+30\,m SiO(5--4) data towards ISOSS23053. The three
  panels show in color scale the integrated intensity, 1st and 2nd
  moment maps (intensity-weighted peak velocities and velocity
  dispersions), respectively. These maps were produced by clipping all
  data below an approximate $4\sigma$ threshold of
  20\,mJy\,beam$^{-1}$. The contours show the 1.3\,mm continuum
  emission in $3\sigma$ steps ($1\sigma \sim
  0.16$\,mJy\,beam$^{-1})$. The beam is shown in the bottom-right of
  all panels, and a linear scale bar is shown in the middle
  panel. The cores are labeled in the left panel. The arrow in the
  middle panel marks the position-velocity cut presented in
  Fig.~\ref{pv_sio}.}
\label{23053_sio} 
\end{figure*}

In considering shocks by potential converging gas flows, we also
investigate the SiO(5--4) emission presented in
Fig.~\ref{23053_sio}. Here, SiO is expected to be strongly enhanced by
shocks because Si gets sputtered from the grains and quickly forms SiO
in the gas phase (e.g.,
\citealt{schilke1997a,anderl2013}). Interestingly, the SiO emission is
comparably weak toward the main cores in the region, but it shows
enhanced emission at several positions throughout the cloud. The
strongest integrated SiO emission is identified between the main cores
\#2\&\#3 and \#6. Additional significant SiO emission is found between
the cores \#1 and \#5 as well as in some elongated structures toward
the north of the region (e.g., one labeled as ``stream'' in
Fig.~\ref{23053_sio}). While the peak velocities in the 1st moment map
show the same blue-red dichotomy as the previously shown by the
DCO$^+$ data, the SiO linewidth increase visible in the 2nd moment map
in Fig.~\ref{23053_sio} is very pronounced. We find increased velocity
dispersion again between cores \#2/\#3 and \#6 as well as between the
cores \#1 and \#5 and the northern stream.

\begin{figure}[htb]
\includegraphics[width=0.49\textwidth]{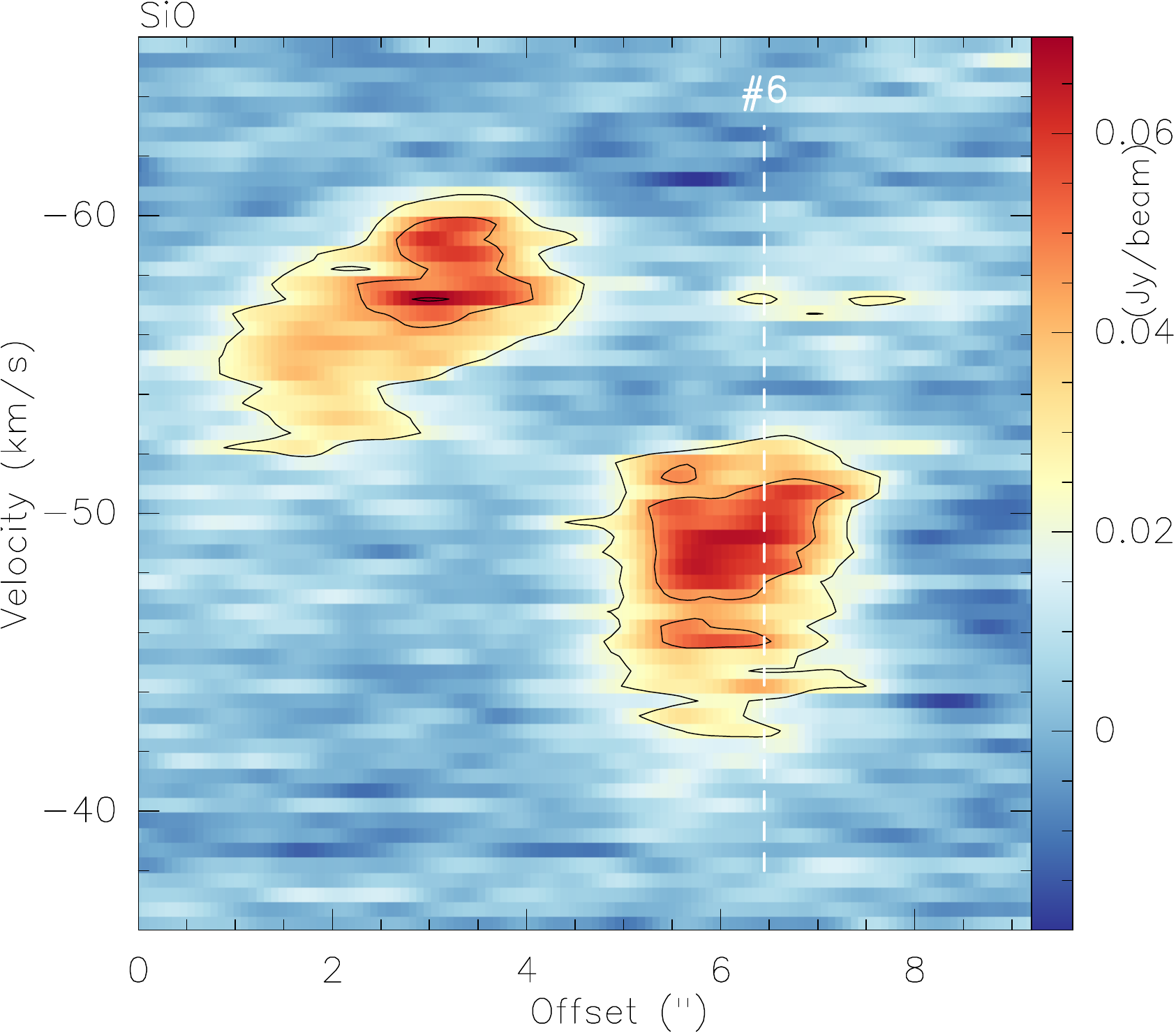}
\caption{Position-velocity diagrams in SiO$(5-4)$ for source
  ISOSS23053 along the cut shown in Fig.~\ref{23053_sio} (middle
  panel). The contours are in $3\sigma$ steps of
  22.5\,mJy\,beam$^{-1}$. The white dashed line marks the position of
  core \#6.}
\label{pv_sio} 
\end{figure}

While SiO is often associated with shocks from outflows (e.g.,
\citealt{chandler2001,palau2006,cabrit2012}), shocks in a more general
sense also from other drivers are sufficient (e.g.,
\citealt{anderl2013}). For example, shocks during cloud formation may
contribute to the SiO formation and emission as well (e.g.,
\citealt{jimenez-serra2010,jimenez-serra2014}). In the case of
ISOSS23053, the stream-like features exhibit elongated structures that
are most likely caused by outflows. The orientation of the
stream-feature labeled in Fig.~\ref{23053_sio} is towards core \#3,
which may be the driver of that outflow structure. However, other
strong SiO emission peaks, in particular those between the cores \#1
and \#5 as well as between cores \#2\&\#3 and \#6, do not necessarily
need to be of protostellar outflow origin. Although core \#1 is
reported to drive an outflow \citep{birkmann2007}, the strong SiO
emission is rather centered between cores \#1 and \#5. All the SiO
features between the cores \#1/\#5 and \#2/\#3/\#6 are spatially
closely linked to the velocity jump best visible in
Fig.~\ref{23053_dco+} or nearby core \#6 in the SiO emission in
Fig.~\ref{23053_sio}. Conducting a position velocity cut approximately
through core \#6 (middle panel of Fig.~\ref{23053_sio}), the
corresponding position-velocity diagram is presented in
Fig.~\ref{pv_sio}. The SiO emission along this cut covers a much
broader velocity range than the DCO$^+$ emission shown in
Fig.~\ref{23053_slices}. Furthermore, the two velocity components
appear even more clearly separated in space.

While it cannot be entirely excluded that these velocity differences
are of protostellar outflow origin, the clear spatial separation of
the blue- and red-shifted emission without a clear driving source at
its center make the outflow origin a less likely scenario. Quantifying
the velocity jump in Fig.~\ref{pv_sio} from the two peak positions in
the position velocity diagram at $3.0''$/-57.1\,km\,s$^{-1}$ and
$6.1''$/-49.2\,km\,s$^{-1}$, we derive an estimate for the velocity
gradient of $\sim$122\,km\,s$^{-1}$\,pc$^{-1}$, more than twice the
value found in the DCO$^+$ emission. These velocity jumps are
discussed in Sect. \ref{dynamics} in the context of potential
converging gas flows.

\begin{figure}[htb]
\includegraphics[width=0.49\textwidth]{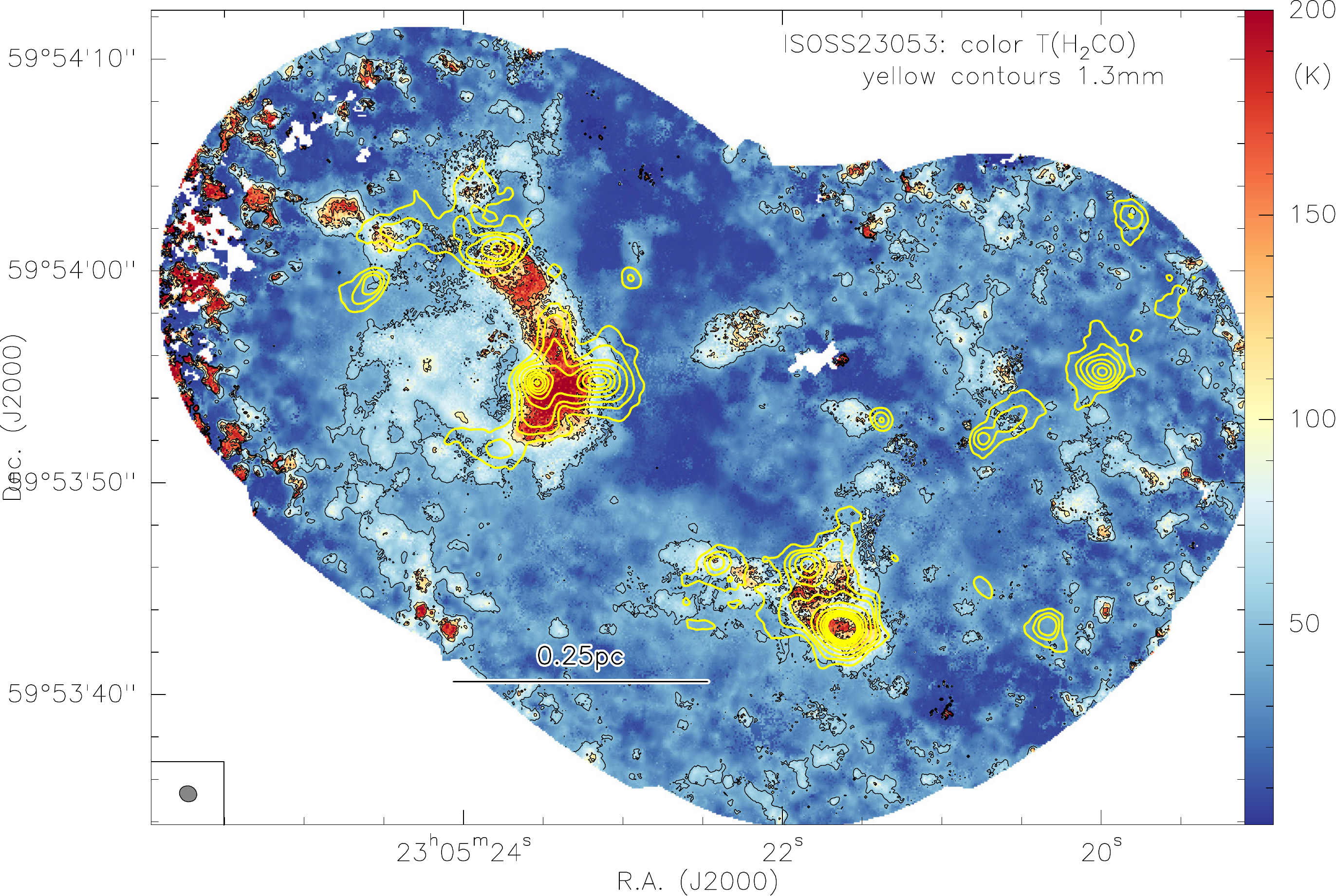}
\includegraphics[width=0.49\textwidth]{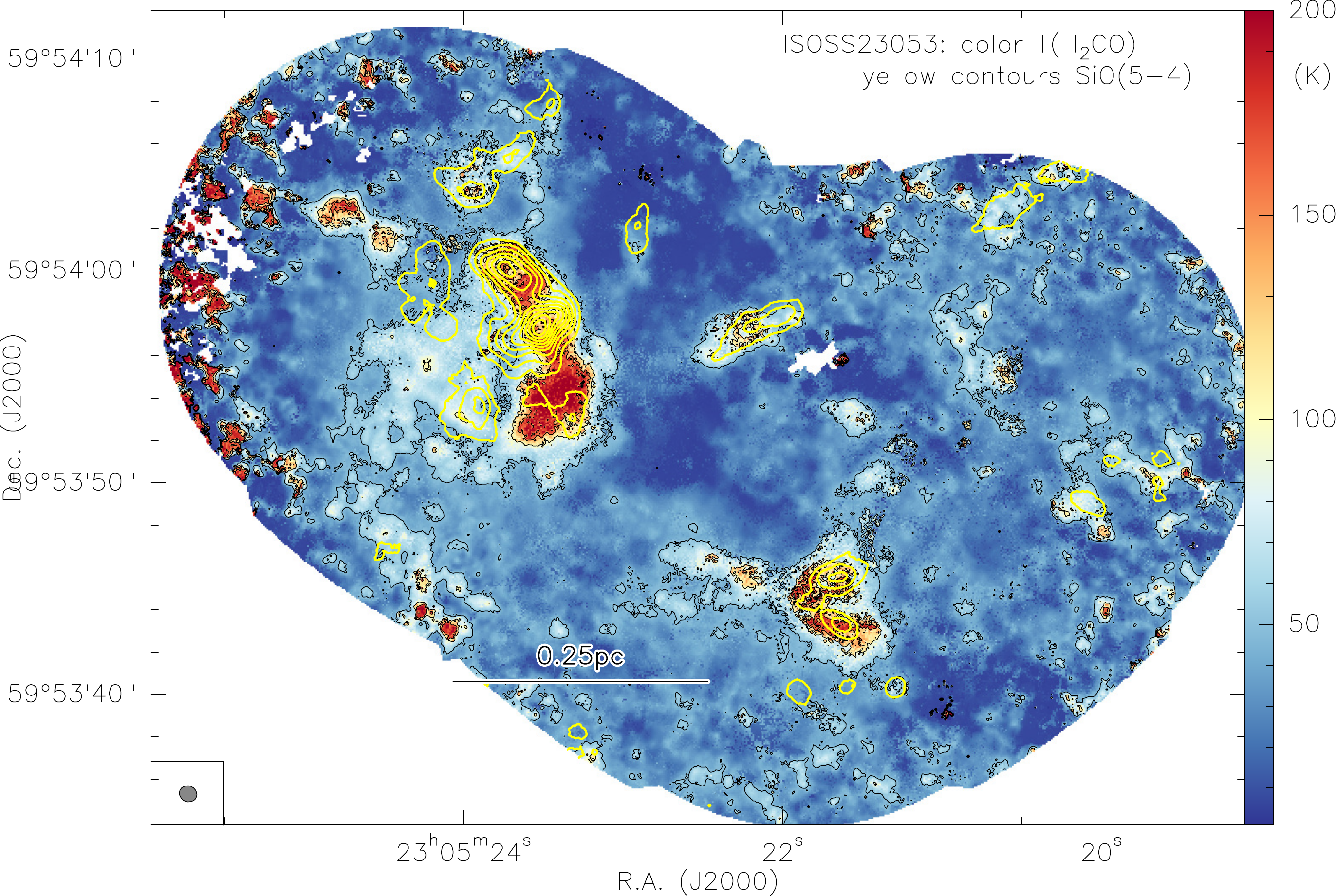}
\caption{Temperature map for ISOSS23053 derived from the H$_2$CO line
  emission. Both panels show, in color scale, the derived temperatures
  with black contour levels of 50, 100, and 150\,K.  The high
  temperatures in particular at the eastern edge of the map are not
  real but are caused by the lower signal-to-noise ratio there. The
  yellow contours in the top and bottom panels present the 1.3\,mm
  continuum and the integrated SiO(5--4) emission, respectively. The
  1.3\,mm continuum contours are in $3\sigma$ steps ($1\sigma \sim
  0.16$\,mJy\,beam$^{-1})$. The SiO emission is integrated from -61 to
  -43\,km\,s$^{-1}$ clipping all data below an approximate $4\sigma$
  threshold of 20\,mJy\,beam$^{-1}$. Contour levels are from 10 to
  90\% (step 10\%) of the integrated peak emission of
  1.58\,Jy\,beam$^{-1}$\,km\,s$^{-1}$. The beam is shown in the
  bottom-left of each panel, and a linear scale bar is shown as well.}
\label{t_23053} 
\end{figure}

The temperature structure in ISOSS23053 shows a different distribution
compared to what we saw before for ISOSS22478. Figure \ref{t_23053}
shows the gas temperature derived again from the H$_2$CO emission
lines in comparison with the 1.3\,mm continuum (top panel) and the
integrated SiO emission (bottom panel). While we still find
significantly large areas with temperatures around 10\,K (dark blue in
Fig.~\ref{t_23053}), similarly large areas are already above 30\,K
(lighter blue). However, even more importantly, ISOSS23053 exhibits, in
significant parts of the map, temperatures above 50\,K and reaching
even 100\,K and above. If we compare these areas of enhanced
temperatures with the dust continuum emission, that traces the dense
cores, and with the SiO emission, that should trace the shocked gas,
we clearly find that the high-temperature gas is preferentially
associated with the shocks. This is not only the case for the areas
between the main cores, but we also find this temperature increases 
toward the stream-like SiO structures that likely trace outflowing
gas. However, we stress that the temperature enhancements as well as
shock-tracing SiO emission are both found toward the region of the
strong velocity jump in this regions. All these physical
characteristics appear to be closely linked with what may happen in
the case of converging gas flows (Sect. \ref{dynamics}).

\section{Discussion}
\label{discussion}

\subsection{Fragmentation}
\label{fragmentation}

Comparing the estimated masses and column densities to the
corresponding values of the original CORE study of 20 high-mass
star-forming regions in the more evolved evolutionary stages of
high-mass protostellar objects (HMPOs), massive young stellar objects
(mYSOs), and ultracompact H{\sc ii} regions \citep{beuther2018b}, we
find that the masses are covering a similar range. In contrast to
that, the column densities of the original CORE sources are between
$\sim$10$^{23}$ and $\sim$10$^{25}$\,cm$^{-2}$, about one order of
magnitude larger than what we find here for the two younger
regions. While this may be attributed to the younger evolutionary
stage of the ISOSS sources presented here, we caution that the spatial
resolution of the original CORE data was about a factor of 2 better
(because of uniform weighting there and natural weighting here, see
Sect. \ref{data}). Hence, with the higher spatial resolution, it was
easier to resolve the highest column density peaks for the more
evolved sources from the original CORE sample.

The mass-size relation shown in Fig.~\ref{relations} is a bit puzzling
because of the a priori uncertainty in the temperature
estimates. While the $T_{\rm 20K}$ approach results in a mass-size
relation with $M\propto r^{3.0\pm 0.2}$, which would correspond to
constant mean volume densities for the sampled cores around
$10^6$\,cm$^{-3}$, the $T_{\rm H2CO}$ approach results in a flatter
mass-size relation $M\propto r^{2.0\pm 0.3}$. The latter agrees
closely with the slope expected from Larson's 3rd relation (e.g.,
\citealt{ballesteros2020}) and corresponds to constant mean column
densities. In the literature, mass-size relations with various
exponents can be found. While some studies report relations close to the
classical $r^{-2.0}$ (e.g.,
\citealt{heyer2009,lombardi2010,ballesteros2019}), steeper relations
have been found as well. For example, \citet{kainulainen2011} found a
mass-size relation with exponent 2.7 when sampling only the highest
column density parts of their studied molecular clouds (those regions
where the column density probability density functions leave the
lognormal shape and rather flatten out). It is interesting to note
that also in the recent comparison between the extreme environments of
the nuclear starburst of the extragalactic system NGC253 and the
central molecular zone of our Milky Way mass-size relations consistent
with $M\propto r^3$ were found \citep{krieger2020}.

\citet{ballesteros2012,ballesteros2019,ballesteros2020} discuss in
detail the notion that mass-size relations with power-law indices from below 2 up
to 3 have been reported in the literature (e.g.,
\citealt{mookerjea2004,lada2008,roman-zuniga2010,kainulainen2011,konyves2015,zhang2017,veltchev2018}). They
reiterate that mass-size relations derived from constant column
density thresholds (which is approximately the case for the $5\sigma$
intensity thresholds used here in the core finding algorithm), if the
filling factor of the densest portions of the cores is small, should
necessarily exhibit constant mean column densities, and thus, a
$M\propto r^{2.0}$ relationship. However, if the small filling-factor
hypothesis for the densest gas in the continuum data and core
identification were not fulfilled (e.g., Fig.~\ref{cont_num}), the
cores may exhibit sharp boundaries and the lower column density
material would not contribute substantially to the mean column
density. Such observational conditions would imply that the column
density probability density functions (PDFs) do not fall as fast as
necessary to produce a $M\propto r^{2.0}$ slope
\citep{ballesteros2012}. Setting this picture into context with our
observations, we cannot claim with certainty which temperature
assumption is better for all cores and hence which mass-size relation,
corresponding either to constant mean column density or to constant
mean volume density. Possibly, some intermediate slope may represent
the regions best. Hence, while our data are consistent with the
classical 3rd Larson relation of constant mean column densities, some
steeper slope, that results from not well sampled column density PDFs
of the cores, cannot be excluded.

\begin{figure}[htb]
\includegraphics[width=0.49\textwidth]{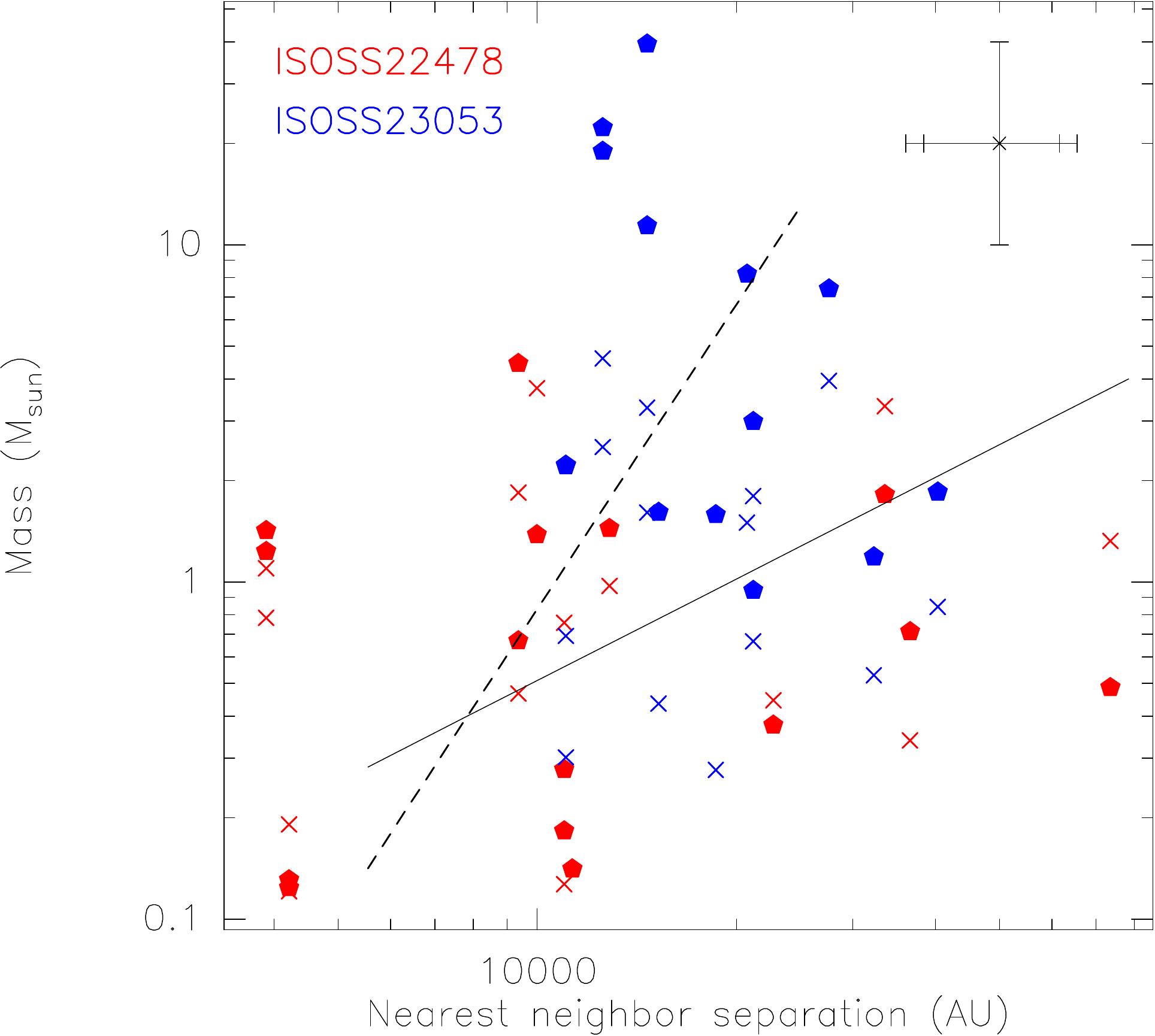}
\caption{Core masses against nearest-neighbor separations from the
  minimum spanning tree analysis. The full pentagons show the results
  estimated assuming constant temperatures ($T_{\rm 20K}$) whereas the
  crosses present the data assuming the temperatures derived from the
  H$_2$CO line data ($T_{\rm H2CO}$). The full line corresponds to the
  Jeans lengths and Jeans masses calculated at 20\,K for a density
  grid between $5\times 10^3$ and $10^6$\,cm$^{-3}$. For comparison,
  the dashed line corresponds to the Jeans lengths and Jeans masses
  calculated at a fixed density of $5\times 10^5$\,cm$^{-3}$ with
  temperatures between 10 and 200\,K. The error-bars in the top-right
  correspond to a factor 2 uncertainty for masses and 1\,kpc distance
  uncertainty for the nearest-neighbor separations (resulting in the
  different error-bar end markers for the separations).}
\label{sep_mass} 
\end{figure} 

We can also set the derived masses and the projected nearest-neighbor
separations into context within the classical Jeans analysis by
estimating the Jeans masses and Jeans lengths for the typical parental
gas clumps. Assuming mean volume densities of the original gas clump
of $\sim 10^5$\,cm$^{-3}$ at temperatures of $\sim$20\,K (e.g.,
\citealt{hennemann2008}), the corresponding Jeans length and Jeans
mass (e.g., \citealt{stahler2005}) are $\sim$17500\,au and
$\sim$0.9\,M$_{\odot}$, respectively. Both of these values are
approximately in the regime of the majority of core masses and
separations (Table \ref{cont_par} and
Fig.~\ref{global_sep}). Estimating the mean nearest-neighbor
separations, we get values of $\sim$17170\,au and $\sim$19560\,AU for
ISOSS22478 and ISOSS23053, respectively. These values also agree
roughly with the estimated Jeans length, similarly to what was found by
\citet{sanhueza2019}.

To show this global trend in more detail for the individual cores,
Figure~\ref{sep_mass} presents the core masses versus the nearest
neighbor separations, where the latter are used as proxies for the
Jeans lengths. Corresponding relations for the Jeans analysis are
shown there as well. While there is indeed scatter within the plot,
the scatter is rather uniformly distributed around the typical Jeans
relations. The scatter is slightly smaller for the masses derived with
the $T_{\rm H2CO}$ approach. Following \citet{sanhueza2019}, we can
correct the nearest neighbor separations for average projection
effects with an average correction factor of $\pi/2\sim 1.57$. This
comparably small factor does not significantly affect the general
correspondence of the Jeans relations and the data in
Fig.~\ref{sep_mass}.

This is again different to the original CORE sample of more evolved
sources where the derived parameters scattered above the Jeans
relations (Fig.~13 in \citealt{beuther2018b}). In that context,
\citet{beuther2018b} considered what would happen if in
the Jeans analysis, we replaced  the thermal sound speed by the turbulent velocity
dispersion. The expected relations would shift to the top-right in the
mass-size relation, not improving the congruence between Jeans
analysis and data. Furthermore, as visible in Figs.~\ref{spec_22478}
and \ref{spec_23053}, the observed FWHM of individual spectral
components are pretty low, typically around or even below
1\,km\,s$^{-1}$ (thermal line width of DCO$^+$ at 20\,K $\sim
0.17$\,km\,s$^{-1}$). Hence, turbulent contributions to the line width
are less strong than would be expected in the turbulent core scenario
of high-mass star formation \citep{mckee2003}. Regarding filamentary
fragmentation, in their ALMA study of very young high-mass
star-forming regions \citet{svoboda2019} estimated the cylindrical
fragmentation scale to be roughly a factor of 3.5 of larger than the typical
Jeans length. In that cylindrical framework, the Jeans regimes
outlined in Fig.~\ref{sep_mass} (dashed and continuous lines) would
shift by that factor to the right, almost outside the plot. Hence,
cylindrical fragmentation seems not the most important process in our
regions either.

These considerations further support the idea that thermal Jeans
fragmentation can explain our data and that turbulence as well as
cylindrical fragmentation are unlikely to play a significant role in
the fragmentation of at least these two very young high-mass
star-forming regions. Similar results have also been reported in, for
example, \citet{palau2014,palau2015,palau2018}, \citet{henshaw2017},
\citet{cyganowski2017}, \citet{klaassen2018} as well as in the recent
ALMA studies of very young high-mass star-forming regions by
\citet{svoboda2019} and \citet{sanhueza2019}. However, there are also
reports in the literature that support the notion that for some other
sources, turbulent pressure may be more important (e.g.,
\citealt{wang2011,wang2014,zhang2015,sadaghiani2020}).

\subsection{Gas dynamics}
\label{dynamics}

While the original CORE program focused more on the fragmentation of
the parental gas clumps and the inner massive disks
\citep{beuther2018b}, the CORE-extension program here focuses on the
dynamics of the gas clumps. It is interesting to see that the
observations of the gas dynamics of the two target regions do differ
in many ways (Sect. \ref{kinematics}).

The intermediate-mass region ISOSS22478 exhibits a more
filamentary structure with velocity gradients along and across the
filaments as well as multiple velocity components at the locations of
the cores. Similar kinematic signatures along filamentary low- and
high-mass star-forming regions have been reported in, e.g.,
\citet{fernandez2014}, \citet{henshaw2014}, \citet{beuther2015b} or
\citet{hacar2017}. The temperature distribution of ISOSS22478 exhibits
relatively uniform low temperatures between roughly 10 and
30\,K. Furthermore, we do not find strong signatures of shocks,
neither in SiO emission nor specific temperature enhancements.

In comparison, the more massive region ISOSS23053 does not show
a particularly filamentary structure, however, a few of the cores are
aligned along a narrow spatial strip where two different velocity
components of two separate cloud structures overlap. Position-velocity
slices across that strip clearly reveal two velocity components and a
very steep gradient between the two components of roughly
50\,km\,s$^{-1}$pc$^{-1}$ measured in DCO$^+(3-2)$ and even
$\sim$122\,km\,s$^{-1}$pc$^{-1}$ measured in SiO$(5-4)$ . This
velocity gradient is significantly larger than that reported by
\citet{hacar2017} of 5--7\,km\,s$^{-1}$pc$^{-1}$ for the gravitational
collapse of the OMC-1 molecular cloud. Furthermore, for ISOSS23053,
the region of velocity overlap shows strong SiO emission as well as a
large temperature increase from roughly 20\,K in the environmental
cloud up to 100\,K and more in the velocity overlap region. The
combination of a velocity jump, enhanced SiO emission as well as
temperature increase are all indicative of shocks that may have been
caused by converging or colliding gas flows.

These two regions may at first sight appear to represent two
potentially different modes of cloud formation. However, the different
signatures in the data could occur based on similar large-scale cloud
collapse processes. Multiple velocity components along filamentary
structures have regularly been observed in various star-forming
regions of different masses (e.g.,
\citealt{hacar2013,hacar2018,kirk2013,fernandez2014,henshaw2014,tackenberg2014,beuther2015b,dhabal2018}),
but there are also significant observational differences reported. For
example, some regions of low and high mass reveal sub-filaments (or
fibers) that may collide and form larger-scale filamentary structures
during the formation process (e.g.,
\citealt{hacar2013,hacar2017b,hacar2018,henshaw2014,smith2014}). Other
observations have revealed velocity gradients across filaments (e.g.,
\citealt{palmeirim2013,fernandez2014,beuther2015b,shimajiri2019}) that
may be explained by gravity-induced accretion from a magnetized
sheet-like molecular cloud (e.g.,
\citealt{chen2015b,chen2020}). However, we should keep in mind that in
these clouds, often not just one signature exists but that several
signatures occur within the same region. For example, studies of
Serpens south as well as of IRDC\,18223 reveal velocity gradients
perpendicular to the clouds in some part of the filaments, but in
other parts of the clouds these studies find clearly distinct multiple
velocity components that may resemble more the sub-filament signatures
\citep{fernandez2014,dhabal2018,beuther2015b}. \citet{chen2020} point
out that multiple velocity-components along a filament do not
necessarily always have to be real structures but that the
observational bias that different spectral lines trace different
density regimes can also create artificial multiple velocity
components in observations (see also
\citealt{ballesteros2002,zamora-aviles2017}). Different inclination
angles and optical depth effects may also induce a similar bias in
some data as well.

From a physical point of view, both scenarios, namely, colliding pre-existing
filaments and gravity-induced accretion in sheet-like structures have
a feature in common: both basically rely on the large-scale gravitational
collapse of the molecular clouds. Low magnetization may result in more
early fragmentation and then interacting pre-existing sub-filaments
whereas higher magnetization should result in smoother structures and
more coherent accretion onto filaments (see for example,
\citealt{hennebelle2019}, and references there in).

The distinct signature of a steep velocity gradient or even sharp
velocity jump in ISOSS23053 can be interpreted as a signature of
converging and colliding gas flows. For example, the cloud formation
models with colliding flow conditions by \citet{gomez2014} produce
similar velocity gradient structures, as seen in ISOSS23053. However,
other simulations of cloud collapse, either hydrodynamic or
magneto-hydrodynamic, can produce qualitatively similar velocity
gradients as well (e.g., \citealt{smith2013,chen2020}). The
outstanding aspect of the ISOSS23053 case is the steepness of the
gradient with $\geq$50\,km\,s$^{-1}$pc$^{-1}$. While variations in
inclination angles can induce differences in the slope of such a
gradient, it is also possible that different physical properties,
for instance, different magnetization or external compression may enhance
the observed gradients. It appears that the smaller and smoother
velocity gradients, as observed, for example, in the Serpens south,
IRDC\,18223, or the Orion integral shape filament
\citep{fernandez2014,beuther2015b,hacar2017}, can be explained mainly
by gravitational collapse, whereas the sharp velocity jump observed
here in ISOSS23053 may require an external cause that drives the
converging gas flow. Gravitational collapse can then occur at the
colliding flow interface.

Differentiating these variations requires additional observational as
well as theoretical work. For example, it will be important to infer
the magnetic field for a sample of clouds and investigate whether
steep velocity gradients or multiple velocity components may correlate
with the strength or orientation of the environmental magnetic
fields. One obvious difference between the two regions is their
Galactic latitude. While ISOSS23053 is close to the Galactic plane
with a latitude of $-0.28$\,deg, ISOSS22478 is located at larger
Galactic latitudes of $+4.26$\,deg. It is reasonable that closer to
the Galactic plane, along with more star formation activity, there is
also a more dynamic environment that can further amplify the potential
dynamic gas structure and velocity gradients. On the simulation side,
we would need to sample a broader range of magnetizations to infer
whether stronger magnetic fields indeed favor smoother accretion onto
filamentary structures, while lower magnetization may favor early
formation of sub-filaments that could then collide during ongoing
collapse.

What all of the above scenarios have in common is that they are based on a
dynamical cloud collapse scenario that forms filamentary structures
during the collapse motions. Environmental effects such as external
shocks, converging gas flows or different magnetizations come into
play, but they are all part of a dynamical collapse and fragmentation
scenario.

\section{Conclusions and summary}
\label{conclusion}

In the context of the NOEMA large program CORE, which studies, among
other topics, the fragmentation processes during high-mass star
formation, we studied two young regions of intermediate- to high-mass
star formation, namely ISOSS22487 and ISOSS23053. We used NOEMA and
the IRAM 30\,m telescope to produce high spatial resolution
($\sim$0.8$''$) mosaics to investigate their larger-scale
fragmentation and cloud formation processes. In both regions, many
fragments can be identified in the 1.3\,mm continuum emission, and a
diversity of spectral lines can be imaged over the entire field of
view. We concentrate on the fragmentation based on the continuum
emission, and on the kinematics of the clouds based on the DCO$^+$,
H$_2$CO, and SiO emission. A forthcoming paper will present the entire
spectral line data and a chemical analysis (Gieser et al.~in prep.).

The 1.3\,mm continuum data reveal 15 and 14 cores in ISOSS22478 and
ISOSS23053, respectively. Most of them are associated with filamentary
sub-structures in the cloud.  Depending on the assumed temperatures of
the cores, we estimate slightly different core masses. These
differences then result in mass-size relations varying possibly
between $M\propto r^2$ and $M\propto r^3$. Hence, while the data are
consistent with the classical mass-size relation of $r^2$ expected
from the 3rd Larson relation for regions with constant mean column
densities, a steeper relation cannot be ruled out. The latter could be
explained because the cores in these maps have sharp boundaries and,
thus, the lower iso-contours in the mm emission do not have much
larger areas than higher iso-contours. Such steeper mass-size relation
would correspond to similar mean densities for the cores. Furthermore,
the correlation of the core masses with their nearest-neighbor
separations is consistent with thermal Jeans fragmentation. In
contrast to the original CORE sample of slightly more evolved sources,
here, we barely find separations at our spatial resolution
limit. Hence, our data resolve the large-scale fragmentation of the
parental gas cloud well.

The kinematic analysis reveals very different observational signatures
between the two regions. While ISOSS22478 is more filamentary in nature
with several velocity components along the length of the filament,
ISOSS23053 shows at the locations of the most massive cores a very
steep velocity jump between two velocity components. The velocity
gradient is roughly 50\,km\,s$^{-1}$pc$^{-1}$ measured in DCO$^+(3-2)$
and even $\sim$122\,km\,s$^{-1}$pc$^{-1}$ measured in
SiO$(5-4)$. While these signatures appear disjointed at first sight, we
argue that all these kinematic observations can be understood within the
framework of dynamically collapsing clouds. Depending on the
environmental properties, for example, external shocks causing
converging gas flows or magnetized sheets that preferentially form
filaments that accrete gas from the parental gas structure, different
observational signatures of a global cloud collapse may be
present. Further observational and theoretical investigations are
needed to pin down in more detail how much the physical processes
(e.g., magnetization, external compression) as well as the
observational biases (e.g., tracers of different density, optical
depth effects, inclination angles) contribute to the overall dynamical
collapse of the clouds.

\begin{acknowledgements} 
  This work is based on observations carried out under project number
  L14AB with the IRAM NOEMA Interferometer and the IRAM 30\,m
  telescope. IRAM is supported by INSU/CNRS (France), MPG (Germany)
  and IGN (Spain). We like to thank an anonymous referee for very
    useful comments. H.B.~and S.S.~acknowledges support from the
  European Research Council under the Horizon 2020 Framework Program
  via the ERC Consolidator Grant CSF-648505. H.B. also acknowledges
  support from the Deutsche Forschungsgemeinschaft in the
  Collaborative Research Center (SFB 881) “The Milky Way System”
  (subproject B1). A.P.~acknowledges financial support from CONACyT
  and UNAM-PAPIIT IN113119 grant, M\'exico. DS acknowledges support by
  the Deutsche Forschungsgemeinschaft through SPP 1833: ``Building a
  Habitable Earth'' (SE 1962/6-1). RK acknowledges financial support
  via the Emmy Noether Research Group on Accretion Flows and Feedback
  in Realistic Models of Massive Star Formation funded by the German
  Research Foundation (DFG) under grant no. KU 2849/3-1 and KU
  2849/3-2.
\end{acknowledgements}

%\bibliography{../../bibliography}   
%\bibliography{/Users/henrikbeuther/tex/bibliography}
%\bibliographystyle{aa}    % this does the style, aa.bst necessary
%\input{40106.bbl}

\end{document}